\documentclass[aps,showkeys,showpacs,superscriptaddress,nofootinbib,twocolumn,10pt]{revtex4-1}
\usepackage{graphicx}  
\usepackage{amsmath,amsfonts,bbm,MnSymbol}   
\usepackage{amsfonts}
\usepackage[breaklinks=true,colorlinks=true,linkcolor=blue,urlcolor=blue,citecolor=blue]{hyperref}
\usepackage{comment}
\usepackage{color}
\usepackage[makeroom]{cancel}
\pdfoutput=1

\renewcommand{\[}{\begin{equation}}
\renewcommand{\]}{\end{equation}}
\newcommand{\ket}[1]{|#1\rangle}
\newcommand{\bra}[1]{\langle#1|}

\newcommand{\pro}[2]{|#1\rangle\langle#2|}
\newcommand{\mean}[1]{\langle#1\rangle}
\newcommand{\abs}[1]{|#1|}
\newcommand{\ov}[1]{\overline{#1}}

\newcommand{\norm}[1]{\left\lvert#1\right\rvert}

\newcommand{\A}{\alpha}
\newcommand{\B}{\beta}

\newcommand{\bd}{\boldsymbol{d}}
\newcommand{\bA}{\boldsymbol{\hat{A}}}
\newcommand{\bb}{\boldsymbol{b}}
\newcommand{\bg}{\boldsymbol{\gamma}}

\begin{document}

\title{Optimal probe states for the estimation of Gaussian unitary channels}

\author{Dominik \v{S}afr\'{a}nek}
\email{dominik.safranek@univie.ac.at}
\address{School of Mathematical Sciences, University of Nottingham, University Park,
Nottingham NG7 2RD, United Kingdom}
\address{Faculty of Physics, University of Vienna, Boltzmanngasse 5, 1090 Vienna, Austria}
\author{Ivette Fuentes}\thanks{Previously known as Fuentes-Guridi and Fuentes-Schuller.}
\affiliation{School of Mathematical Sciences, University of Nottingham, University Park,
Nottingham NG7 2RD, United Kingdom}
\affiliation{Faculty of Physics, University of Vienna, Boltzmanngasse 5, 1090 Vienna, Austria}

\date{\today}

\begin{abstract}
We construct a practical method for finding optimal Gaussian probe states for the estimation of parameters encoded by Gaussian unitary channels. This method can be used for finding all optimal probe states, rather than focusing on the performance of specific states as shown in previous studies. As an example, we apply this method to find optimal probes for the channel that combines the phase-change and squeezing channels, and for generalized two-mode squeezing and mode-mixing channels. The method enables a comprehensive study of temperature effects in Gaussian parameter estimation. It has been shown that the precision in parameter estimation using single mode states can be enhanced by increasing the temperature of the probe. We show that not only higher temperature, but also larger temperature differences between modes of a Gaussian probe state can enhance the estimation precision.
\end{abstract}

\pacs{03.67.-a, 06.20.-f, 03.65.Ta}
\keywords{quantum metrology, Gaussian states, local estimation theory}

\maketitle

\section{Introduction}

In recent years, the interest in quantum technologies has increased since this research area is in the brink of reaching the commercialization stage. Important theoretical and experimental efforts are underway to exploit quantum properties, such as squeezing and entanglement, in the development of a new generation of sensors that improve on the precision of their classical counterparts by orders of magnitude~\cite{Giovannetti2004a}. However, there is still substantial work to be done on improving on the capability for preparing certain states, on the protection of states from decoherence, on being able to implement specific measurements and on finding optimal probe states to achieve the highest possible sensitivity. In this paper we develop a practical method for finding optimal probe  Gaussian states, which are a family of states that are very accessible in experiments.

In quantum metrology probe states are quantum states used to optimally estimate an unknown parameter of a quantum channel.  A quantum channel is a transformation that can be unitary or correspond to a complete positive map in the case where the system interacts with the environment. The typical strategy is simple~\cite{Paris2009a}:  The probe state is fed into the channel, the channel encodes the parameter on the state of the system and, finally, measurements are performed with the aim of gaining maximal information about the parameter. Some probe states are affected more by than others for a given channel, i.e., they are more sensitive. Channels of interest in this paper are Gaussian channels, which transform a Gaussian state into another Gaussian state. Finding the optimal family of probe states for a given channel is one of the main tasks of quantum metrology. The aim is to achieve the Heisenberg limit, which is the optimal rate at which the accuracy of a measurement can scale with the energy stored in a probe state.

Gaussian states are usually not optimal probe states. When dephasing is not present non-Gaussian states such as GHZ states usually perform as better probes. However, previous theoretical studies show that Gaussian states can be still effectively used for the estimation of Gaussian channels such as phase changing~\cite{Monras2006a,Aspachs2008a,Sparaciari2015a,sparaciari2016gaussian}, squeezing~\cite{Milburn1994a,Chiribella2006a,Gaiba2009a}, two-mode squeezing and mode-mixing channels~\cite{Gaiba2009a}. Previous studies analyzed specific channels and for each channel \emph{only one} probe state achieving the Heisenberg limit was found. In addition, Gaussian state metrology was often restricted to pure states. Less attention was given to thermal states, which are of great relevance in practice. In the laboratory, quantum states can never be isolated from the environment which thermalises the states. In this paper we develop a formalism that can be effectively used to study \emph{any} Gaussian probe state for \emph{any} one- and two-mode Gaussian unitary channels. Moreover, we develop methods to find \emph{all} optimal Gaussian probe states for these channels. We take advantage of recent developments in the phase-space formalism of Gaussian states~\cite{Pinel2012a,Pinel2013b,Monras2013a,Jiang2014a,Gao2014a,Safranek2015b,Banchi2015a}, making use of Euler's decomposition of symplectic matrices, the Williamson decomposition of the covariance matrix in the complex form, and expression for the quantum Fisher information in terms of the Williamson decomposition~\cite{Safranek2015b}. These techniques enable us to simplify expressions so that formulas can be easily used in practical applications.  As an example, we derive optimal states for channels that, to our knowledge, have not been optimised before. These are the channel combining the phase change and squeezing, and generalized mode-mixing and two-mode squeezing channels. Interestingly, we find that in the estimation of two-mode channels, separable states consisting of two one-mode squeezed states perform as well as their entangled counterpart: two-mode squeezed states. This shows that entanglement between the modes does not enhance precision in this case.

Our formalism also enables us to further our understanding of the effects of temperature in probe states. It has been reported in~\cite{Aspachs2008a} that higher temperature in squeezed thermal states can enhance phase estimation, while higher temperature of displaced thermal states is detrimental. We show that the effects of thermalised probe states on the estimation of Gaussian channels are generic, i.e., for all Gaussian unitary channels, temperature effects are always manifested in multiplicative factors of four types. Two of the factors correspond to the ones previously found in~\cite{Aspachs2008a}. The other two -- newly discovered -- factors show that not only temperature of the probe state, but also temperature difference between different modes of the probe state helps the estimation.


The paper is organized as follows. We first introduce the phase-space formalism for Gaussian states, Gaussian unitary channels, and techniques for the optimal estimation of channel parameters. We present a general framework to find optimal probe states for any Gaussian unitary channel and we study the effects of temperature on the estimation strategy. We apply our formalism to present concrete examples for one- and two-mode Gaussian unitary channels and generalize bounds on the precision of estimation found in~\cite{Milburn1994a,Chiribella2006a,Monras2006a,Aspachs2008a,Gaiba2009a}. In the concluding section we discuss the Heisenberg and the shot-noise limits of our results. Three Appendixes are included providing details on the phase-space description of Gaussian unitary channels (Appendix~\ref{app:derivation_of_S_and_b}), relevant characteristics of the channels (Appendix \ref{app:list_of_sympl_matrices}), and general results for the estimation of two-mode squeezing and mode-mixing channels using a wide class of two-mode probe states (Appendix~\ref{app:full_expressions}).

\section{Quantum metrology on Gaussian states}\label{sec:quantum_metrology_on_GS}

The main aim of quantum metrology is to provide techniques to estimate as precisely as possible a physical parameter encoded in a quantum state. In this section we review techniques that provide lower precision bounds in the estimation of parameters encoded in Gaussian states. This is done conveniently using the phase-space description of \emph{Gaussian states} and \emph{Gaussian unitary operators}.
We consider a system consisting of $N$ Bosonic modes. The operators $\hat{a}_n$ and $\hat{a}_n^\dag$ annihilate and create particles, respectively,  in each mode. In the phase-space description of the system the operators are collected in vector  $\bA:=(\hat{a}_1,\dots,\hat{a}_N,\hat{a}_1^\dag,\dots,\hat{a}_N^\dag)^T$. The commutation relations between the operators can also be written in compact form,
\[\label{def:commutation_relation}
[\bA_{i},\bA_{j}^{\dag}]=K_{ij}\mathrm{id}\quad\!\Rightarrow\quad\! K=
\begin{bmatrix}
I & 0 \\
0 & -I
\end{bmatrix},
\]
where $\mathrm{id}$ denotes the identity element of an algebra and $I$ is the identity matrix. Note that $K^{-1}=K^\dag=K$ and that $K^2=I$.
The \emph{displacement vector} $\bd=(d_1,\dots,d_N,\ov{d}_1,\dots,\ov{d}_N)^T$ and the \emph{covariance matrix} $\sigma$, defined as~\cite{Weedbrook2012a}
\begin{subequations}\label{def:covariance_matrix}
\begin{align}
\bd_i&=\mathrm{tr}\big[\hat{\rho}\boldsymbol{\hat{A}}_i\big],\\
\sigma_{ij}&=\mathrm{tr}\big[\hat{\rho}\,\{\Delta\boldsymbol{\hat{A}}_i,\Delta\boldsymbol{\hat{A}}_j^{\dag}\}\big],
\end{align}
\end{subequations}
correspond to the first and second moments of the field, respectively.
The density operator $\hat{\rho}$ specifies the state of the field and $\{\!\cdot,\cdot\!\}$ denotes the anti-commutator.  The covariance matrix is a positive-definite matrix given in terms of the vector $\Delta\boldsymbol{\hat{A}}:=\boldsymbol{\hat{A}}-\bd$. We emphasise that, to simplify calculations, we choose to use definitions in the complex form, while most authors use the real form.  For more details on their equivalence see Appendix~\ref{app:list_of_sympl_matrices} or~\cite{Arvind1995a,Safranek2015b}.

\emph{Gaussian states} are defined as states that are fully characterized by their first and second moments, while more general states require higher field moments in their description. Gaussian transformations correspond to unitaries $\hat{U}$ that transform Gaussian states into Gaussian states, $\hat{\rho}'=\hat{U}\hat{\rho}\hat{U}^\dag$. These operators are generated via an exponential map with the exponent at most quadratic in the field operators~\cite{Weedbrook2012a},
\[\label{def:Gaussian_unitary}
\hat{U}=\exp\big(\tfrac{i}{2}\bA^\dag W \bA+\bA^\dag K \bg\big),
\]
where $W$ is a Hermitian matrix of the form
\[
W=\begin{bmatrix}
X & Y \\
\ov{Y} & \ov{X}
\end{bmatrix},
\]
$\bg$ is a complex vector of the form $\bg=(\tilde{\bg},\ov{\tilde{\bg}})^T$, and $K$ is the matrix defined in Eq.~\eqref{def:commutation_relation}. In the case that $W=0$, the Gaussian operator~\eqref{def:Gaussian_unitary} corresponds to the \emph{Weyl displacement operator} $\hat{D}(\tilde{\bg})$, while for $\bg=0$ we obtain other Gaussian transformations such as the phase-changing operator, one- and two-mode squeezing operators, or mode-mixing operators depending on the particular structure of $W$. Under the unitary channel~\eqref{def:Gaussian_unitary} the first and the second moments transform according to rule
\[\label{def:transformation}
\bd'=S\bd+\bb,\ \ \sigma'=S\sigma S^\dag,
\]
where, as we prove in Appendix~\ref{app:derivation_of_S_and_b},
\[\label{eq:S_and_b}
S=e^{iKW},\ \
\bb=\Big(\!\int_0^1e^{iKWt}\mathrm{d}t\!\Big)\ \!\bg.
\]
The above identities together with transformation relations~\eqref{def:transformation} are central to this paper. They allow us to transform the  density matrix description of Gaussian states to the phase-space formalism, which is mathematically more convenient.

The matrix $S$ from Eq.~\eqref{eq:S_and_b}, called the \emph{ symplectic matrix}, has the same structure as $W$ and satisfies the relation
\[\label{def:structure_of_S}
S=
\begin{bmatrix}
\A & \B \\
\ov{\B} & \ov{\A}
\end{bmatrix},\ \ SKS^\dag=K.
\]
These two properties define the complex form of the real symplectic group $Sp(2N,\mathbb{R})$. For more details see~\cite{Arvind1995a,Safranek2015b}.

According to the Williamson theorem~\cite{Williamson1936a,deGosson2006a,Simon1998a}, any positive-definite matrix can be diagonalized by the symplectic matrices,
\[\label{def:Williamson_decomposition}
\sigma=SDS^\dag,
\]
where $S$ is the symplectic matrix of the form~\eqref{def:structure_of_S}, and $D$ is the diagonal matrix consisting of the so-called \emph{symplectic eigenvalues}, $D=\mathrm{diag}(\lambda_1,\dots,\lambda_N,\lambda_1,\dots,\lambda_N)$. For the covariance matrix describing a Gaussian state all symplectic eigenvalues are larger than or equal to 1, and a Gaussian state is pure if and only if $\lambda_1=\cdots=\lambda_N=1$.

The Williamson decomposition can be used, for example, to fully parametrize Gaussian states of a given number of modes. Any symplectic matrix~\eqref{def:structure_of_S} can be decomposed using Euler's decomposition as~\cite{Arvind1995a,Weedbrook2012a}
\[\label{def:S_decomposition}
S=
\begin{bmatrix}
U_1 & 0 \\
0 & \ov{U}_1
\end{bmatrix}
\begin{bmatrix}
\cosh{M_{\boldsymbol{r}}} & -\sinh{M_{\boldsymbol{r}}} \\
-\sinh{M_{\boldsymbol{r}}} & \cosh{M_{\boldsymbol{r}}}
\end{bmatrix}
\begin{bmatrix}
U_2 & 0 \\
0 & \ov{U}_2
\end{bmatrix},
\]
where $U_1$ and $U_2$ denote unitary matrices, and $M_{\boldsymbol{r}}=\mathrm{diag}(r_1,\dots,r_N)$ is the diagonal matrix of the squeezing parameters. With a full parametrization of unitary matrices $U_1$ and $U_2$, one can use this decomposition to fully parametrize the covariance matrix via Eq.~\eqref{def:Williamson_decomposition}. Moreover, since the displacement vector is fully parametrized by its elements, we have a full parametrization of Gaussian states. Note, however, that some parameters may not add any additional complexity and can be removed. This is a consequence of the fact that in Eq.~\eqref{def:Williamson_decomposition} some parts of (the decomposition of) $U_2$ vanish, because they commute with the diagonal matrix $\mathrm{diag}(\lambda_1,\dots,\lambda_N)$. We explicitly write the most general one-mode Gaussian state in Sec.~\ref{sec:est_one}, and the most general two-mode Gaussian state in Sec.~\ref{sec:est_two}.

One of the main aims of quantum metrology is to find the ultimate precision limits on the estimation of a physical parameter $\epsilon$ encoded in a quantum state. This is given by the quantum Cram\'er-Rao bound~\cite{BraunsteinCaves1994a,Paris2009a},
\[
\langle (\Delta \hat{\epsilon})^{2}\rangle\geq\frac{1}{M H(\epsilon)},
\]
which gives a lower bound on the mean squared error of the locally unbiased estimator $\hat{\epsilon}$. $M$ denotes the number of measurements taken on identical copies of the state $\hat{\rho}(\epsilon)$, and $H(\epsilon)$ is a quantity called quantum Fisher information. The quantum Fisher information says how precisely we can estimate an unknown parameter $\epsilon$ in a single-shot experiment. For the Williamson decomposition~\eqref{def:Williamson_decomposition} of the covariance matrix of a Gaussian state, the quantum Fisher information reads~\cite{Safranek2015b}
\[\label{def:QFI}
\begin{split}
H(\epsilon)&=\sum_{i,j=1}^N\frac{(\lambda_i-\lambda_j)^2}{\lambda_i\lambda_j-1}\norm{R_{ij}}^2+\frac{(\lambda_i+\lambda_j)^2}{\lambda_i\lambda_j+1}\norm{Q_{ij}}^2\\
&+\sum_{i=1}^N\frac{\dot{\lambda_i}^2}{\lambda_i^2-1}+2\dot{\bd}^\dag\sigma^{-1}\dot{\bd}.
\end{split}
\]
$R$ and $Q$ are submatrices of the matrix $P:=S^{-1}\dot{S}$, satisfying the defining relation of the Lie algebra associated with the symplectic group
\[\label{def:P_1}
P(\epsilon)=
\begin{bmatrix}
R & Q \\
\ov{Q} & \ov{R}
\end{bmatrix},\ \ PK+KP^\dag=0,
\]
and \emph{dot} denotes the derivative with respect to the parameter we want to estimate. For $\epsilon$ such that $\lambda_i(\epsilon)=\lambda_j(\epsilon)=1$, we define the problematic terms in Eq.~\eqref{def:QFI} as $\frac{\dot{\lambda_i}^2}{\lambda_i^2-1}(\epsilon):=\ddot{\lambda_i}(\epsilon)$, and $\frac{(\lambda_i-\lambda_j)^2}{\lambda_i\lambda_j-1}(\epsilon):=0$.

In the estimation of quantum channels we are sometimes interested in the scaling of the quantum Fisher information with the mean number of particles in a probe state. If the quantum Fisher information scales quadratically with $n$, we say the Heisenberg limit is achieved~\cite{Giovannetti2004a}, which signifies a use of quantum resources. In contrast, the linear scaling of the quantum Fisher information is called the shot-noise limit, which can usually be achieved by classical methods.

\section{General framework}\label{sec:gen_framework}

\begin{figure}[t!]
\centering
\includegraphics[width=1\linewidth]{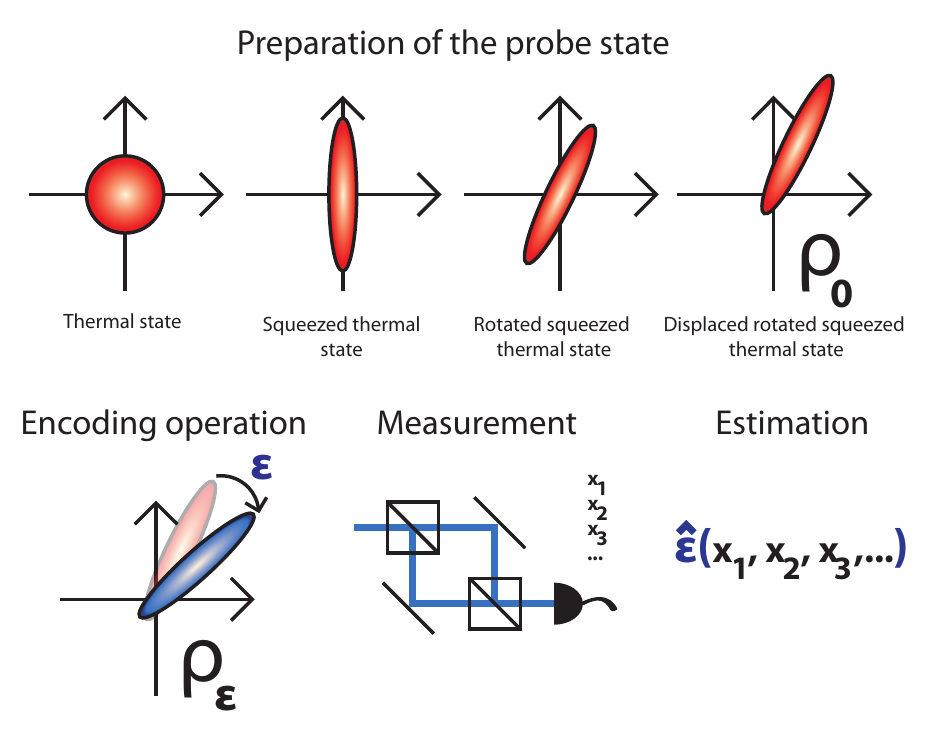}
\caption{Scheme of the usual metrology setting illustrated on a one-mode Gaussian probe state. First, we prepare the state by using various Gaussian operations, then we feed the state into the channel we want to estimate, perform an appropriate measurement, and an estimator $\hat{\epsilon}$ gives us an estimate of the true value of the parameter. In this paper, we are interested in optimizing over the preparation stage for a given encoding Gaussian unitary channel $\hat{U}(\epsilon)$.}\label{fig:scheme}
\end{figure}

In this section we provide a framework for studying the optimal Gaussian probe states for the estimation of Gaussian unitary channels as illustrated in Fig.~\ref{fig:scheme}. Mathematically this is achieved as follows. First, we take a general parametrization of a Gaussian probe state and calculate the quantum Fisher information associated with the channel we estimate. Then we choose parameters of the probe state such that the quantum Fisher information is maximized.

Let us assume we have full control over the preparation of the initial probe state $\hat{\rho}_0\equiv(\bd_0,\sigma_0)$, with the Williamson decomposition $\sigma_0=S_0D_0S_0^\dag$ of the covariance matrix. The diagonal matrix, $D_0$, represents a thermal state and the symplectic matrix $S_0$ together with the displacement vector $\bd_0$ represent operations we are going to perform on this thermal state. After the probe state is created, we feed it into the Gaussian channel that encodes the parameter we want to estimate.

Using Eqs.~\eqref{def:transformation} we find the final state is given by the first and the second moments
\begin{subequations}\label{eq:moments_epsilon}
\begin{align}
\bd_\epsilon&=S_\epsilon\bd_0+\bb_\epsilon,\\
\sigma_\epsilon&=S_\epsilon S_0D_0S_0^\dag S_\epsilon^\dag.
\end{align}
\end{subequations}
As the covariance matrix appears precisely in the form of the Williamson decomposition, we can use formula~\eqref{def:QFI} directly. Applying Eqs.~\eqref{def:structure_of_S}, \eqref{def:P_1}, and \eqref{eq:moments_epsilon}, we derive
\begin{subequations}\label{eq:eq_for_general_method}
\begin{align}
&P=S_0^{-1}P_\epsilon S_0,\label{eq:first_part}\\
&\sum_{k=1}^N\frac{\dot{\lambda_k}^2}{\lambda_k^2-1}=0,\label{eq:second_part}\\
&2\dot{\bd}^\dag\sigma^{-1}\dot{\bd}=2(P_\epsilon\bd_0+S_\epsilon^{-1}\dot{\bb}_\epsilon)^\dag\sigma_0^{-1}(P_\epsilon\bd_0+S_\epsilon^{-1}\dot{\bb}_\epsilon),\label{eq:displacement_part}
\end{align}
\end{subequations}
where we have denoted $P_\epsilon:=S_\epsilon^{-1}\dot{S}_\epsilon$. Due to the unitarity of the channel the symplectic eigenvalues do not change, and expression~\eqref{eq:second_part} vanishes. This scheme can be used for any Gaussian unitary channel. However, in next sections we are going to study Gaussian unitary channels which form a one-parameter unitary group,
\[\label{eq:U}
\hat{U}_\epsilon=\exp\big((\tfrac{i}{2}\bA^\dag W \bA+\bA^\dag K \bg)\epsilon\big),
\]
where $W$ and $\bg$ are independent of $\epsilon$. Because an element of such group is constructed by a substitution in Eq.~\eqref{def:Gaussian_unitary},
\[\label{eq:substitution}
\tfrac{i}{2}\bA^\dag W \bA+\bA^\dag K \bg\rightarrow (\tfrac{i}{2}\bA^\dag W \bA+\bA^\dag K \bg)\epsilon,
\]
we can use Eq.~\eqref{eq:S_and_b} to derive $P_\epsilon=iKW$ and $\dot{\bb}_\epsilon=S_\epsilon\bg$. Inserting these expressions into Eqs.~\eqref{eq:eq_for_general_method} it becomes clear that the resulting quantum Fisher information~\eqref{def:QFI} is independent of $\epsilon$. Given a constant matrix $W$ and a constant vector $\bg$ representing a Gaussian unitary channel, the problem of finding optimal probe states then reduces to finding parameters of the probe state, $S_0,D_0,\bd_0$, such that the quantum Fisher information is maximized. In next sections we will study channels with purely quadratic generators which are characterized by $\bg=\boldsymbol{0}$ in Eq.~\eqref{eq:U}.

\section{Effects of temperature}\label{sec:effects_of_temperature}

It is interesting to note that the symplectic eigenvalues in Eq.~\eqref{def:QFI} appear only in a form of multiplicative factors, independent of other parameters and channels we estimate.

This is particularly interesting from a physical point of view because the symplectic eigenvalues encode temperature. The symplectic eigenvalue describing a thermal state of the harmonic oscillator with frequency $\omega_k$ is given by $\lambda_k=\coth(\frac{\omega_k\hbar}{2kT})$, or alternatively, $\lambda_k=1+2n_{{\mathrm{th}}k}$ where $n_{{\mathrm{th}}k}$ denotes the mean number of thermal bosons in each mode.

In Eq.~\eqref{def:QFI} we can identify four types of multiplicative factors given by symplectic eigenvalues, $\frac{\lambda_k^2}{1+\lambda_k^2}$, $\frac{(\lambda_k+\lambda_l)^2}{\lambda_k\lambda_l+1}$, $\frac{(\lambda_k-\lambda_l)^2}{\lambda_k\lambda_l-1}$, and $\frac{1}{\lambda_k}$.\footnote{We do not count $\frac{1}{\lambda_k^2-1}$ because $\lim_{\lambda_k\rightarrow 1}\frac{1}{\lambda_k^2-1}=+\infty$ while $\lim_{\lambda_k\rightarrow 1}\frac{\dot{\lambda_k}^2}{\lambda_k^2-1}=\ddot{\lambda_k}$. Therefore $\frac{1}{\lambda_k^2-1}$ does not represent a freestanding factor.} First, let us focus on effects of temperature given by the first three types of factors which multiply matrices $R$ and $Q$. These represents sensitivity of squeezing and orientation of squeezing of the probe state with respect to the channel we estimate.
The first type of factor, $\frac{\lambda_k^2}{1+\lambda_k^2}$, is one of the two to appear for the \emph{isothermal} (sometimes called \emph{isotropic}) states for which all symplectic eigenvalues are equal. This class also encompasses all pure states. Because $1\leq\lambda_k\leq+\infty$, we have $\frac{1}{2}\leq\frac{\lambda_k^2}{1+\lambda_k^2}\leq1$, where the lower bound is attained by pure states and the upper bound by thermal states with infinite temperature. This means that for isothermal states temperature helps the estimation with maximal enhancement of a factor of 2, a fact already noted in~\cite{Aspachs2008a}. Next, for mixed multi-mode states we have the second and third type of factors, $\frac{(\lambda_k-\lambda_l)^2}{\lambda_k\lambda_l-1}$ and $\frac{(\lambda_k+\lambda_l)^2}{\lambda_k\lambda_l+1}$. These terms become especially important when there is a large difference between the symplectic eigenvalues. Considering $\lambda_l \rightarrow 1$ we have
\begin{subequations}
\begin{align}
\frac{(\lambda_k-\lambda_l)^2}{\lambda_k\lambda_l-1}&\longrightarrow\lambda_k-1=2n_{{\mathrm{th}}k},\\
\frac{(\lambda_k+\lambda_l)^2}{\lambda_k\lambda_l+1}&\longrightarrow\lambda_k+1=2(n_{{\mathrm{th}}l}+1).
\end{align}
\end{subequations}
Generally, assuming $\lambda_k\gg \lambda_l$ yields
\[
\frac{(\lambda_k-\lambda_l)^2}{\lambda_k\lambda_l-1}\approx\frac{(\lambda_k+\lambda_l)^2}{\lambda_k\lambda_l+1}\approx\frac{2n_{{\mathrm{th}}k}}{2n_{{\mathrm{th}}l}+1}.
\]
This shows that the enhancement by temperature difference is no longer bounded by some fixed value as in the previous case.

If we keep one mode sufficiently cool and the other hot, or if one mode has a high frequency and the other a low frequency, we can, in principle, achieve an infinite enhancement in the estimation of the unknown parameter. 
In general, states with a large variance in energy, which in this case is in the form of thermal fluctuations, have a higher ability to carry information, and thus can carry more information about the parameter we want to estimate. We will refer to this phenomenon as temperature-enhanced estimation.

We have shown that temperature and temperature difference enhances the first two terms in Eq.~\eqref{def:QFI} due to the first three types of factors. However, the opposite behavior is observed in the last term. This last term shows how sensitive the displacement is to the small changes in the parameter of the channel. Factors of the fourth type, $\frac{1}{\lambda_k}$, are hidden in the inverse of the initial covariance matrix in this last term as shown in Eq.~\eqref{eq:displacement_part}. As temperature rises and the symplectic eigenvalues grow to infinity, this factor goes to zero and the precision in estimation diminishes.

\begin{figure}[t!]
\centering
\includegraphics[width=0.49\linewidth]{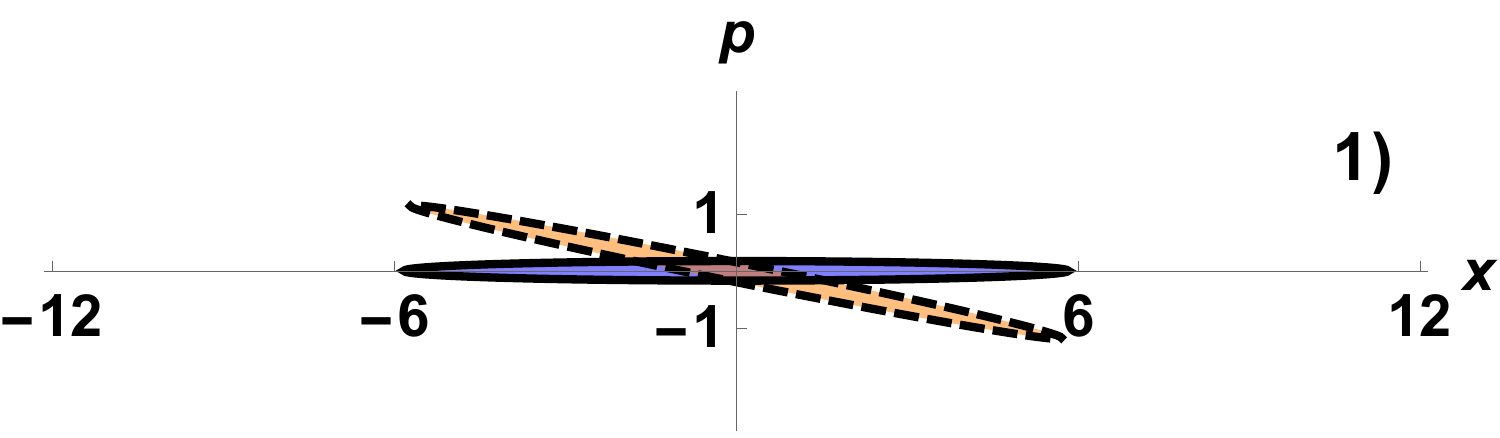}
\includegraphics[width=0.49\linewidth]{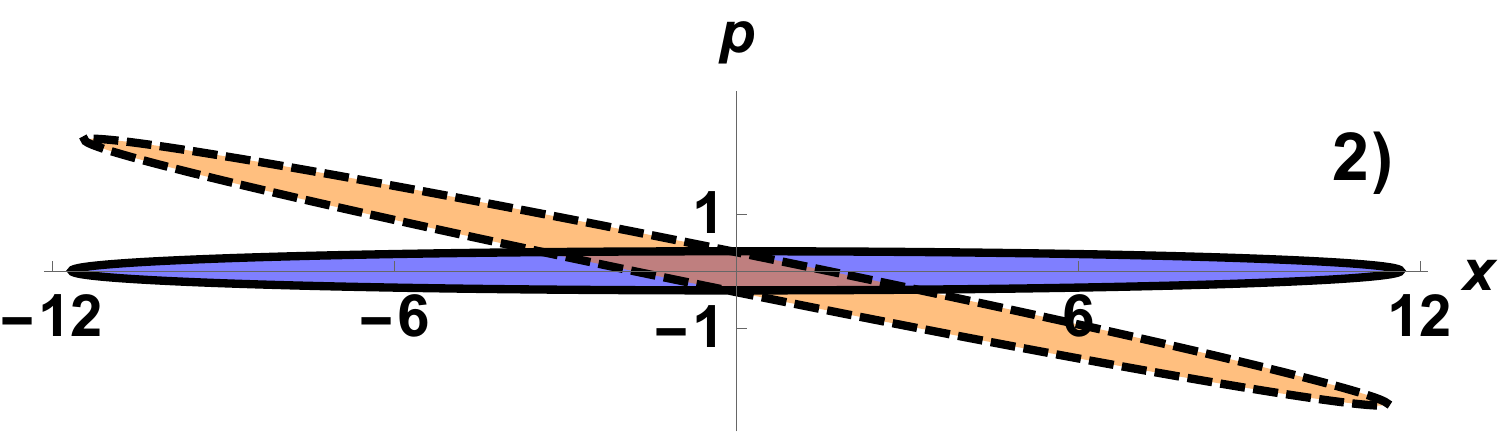}\\
\vspace{0.5cm}
\includegraphics[width=0.49\linewidth]{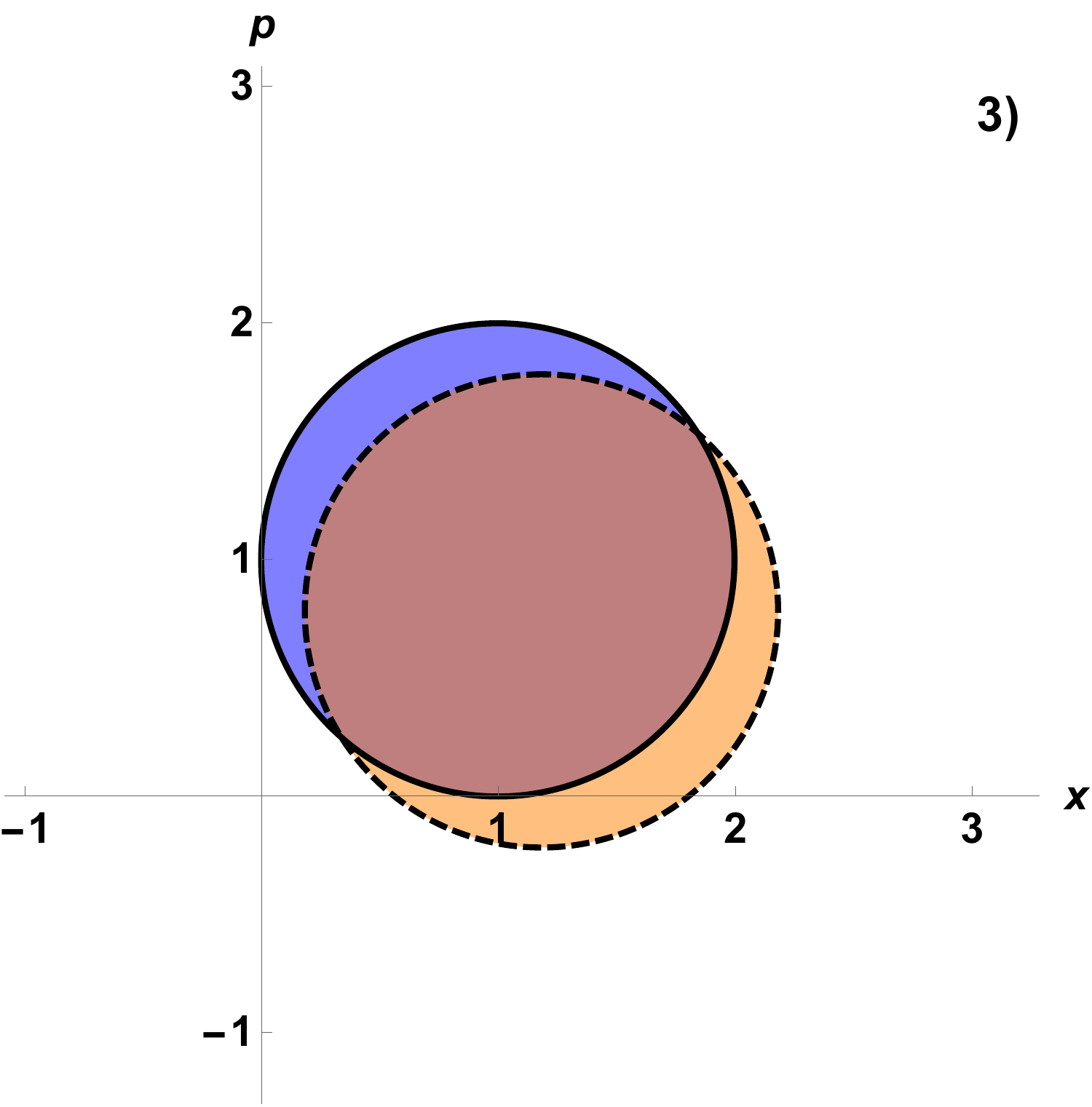}
\includegraphics[width=0.49\linewidth]{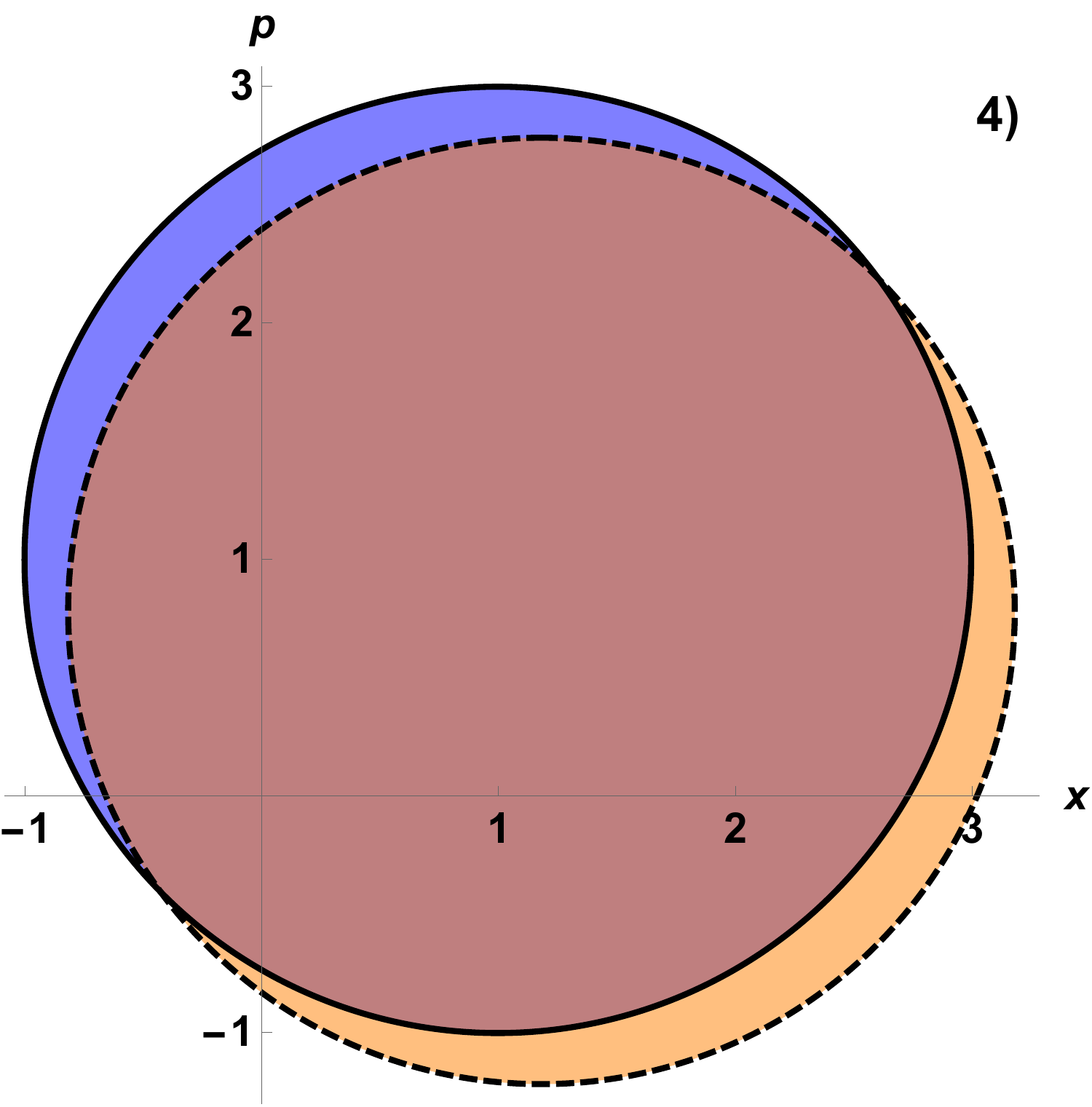}
\caption{Estimation of the phase-changing channel $\hat{R}(\epsilon)$ around point $\epsilon=0$ using various one-mode Gaussian probe states parametrized by Eq.~\eqref{eq:general_1mode_state}. 1) squeezed vacuum [${r=-0.88,\abs{d}=0,\lambda_1=1, H(\epsilon)\approx 16}$], 2) squeezed thermal state [${r=-0.88,\abs{d}=0,\lambda_1=2, H(\epsilon)\approx 25}$], 3) displaced vacuum [${r=0,\abs{d}=1,\lambda_1=1, H(\epsilon)=4}$], and 4) displaced thermal state [${r=0,\abs{d}=1,\lambda_1=2, H(\epsilon)=2}$]. 
The squeezing parameter $r=-0.88$ was chosen in such a way that the squeezed vacuum and the displaced vacuum have the same mean energy $n=1$. We plot covariance matrices in the real form phase-space, before(blue with full line) and after(orange with dashed line) the phase-change $\hat{R}(0.2)$ has been applied. The quantum Fisher information has been calculated from Eq.~\eqref{eq:phase_changing_channel}. The relative overlap of covariance matrices of squeezed states is the same in both cases $1)$ and $2)$, however, the covariance matrix being larger in 2) allows for a better precision in estimation given by the factor $\frac{\lambda_1^2}{1+\lambda_1^2}$. Thermal fluctuations in a squeezed state help the estimation. In contrast, the relative overlap of covariance matrices of displaced states is considerably larger in 4) as compared to 3). Higher thermal fluctuations in a displaced state is detrimental for the estimation. This decrease in precision is given by the factor $\frac{1}{\lambda_1}$.}\label{fig:thermal_phase}
\end{figure}

Let us look at what these factors mean physically for different probe states. Channels quadratic in the field operators, which are given by $\bg=0$, do not affect the displacement of non-displaced probe states such as squeezed thermal states. This means that the precision in estimation of such channels when using non-displaced states will be affected only by factors of the first three types. When using a squeezed thermal state as a probe, temperature and temperature difference in different modes of this probe will always help the estimation. In contrast, when a displaced thermal state is used as a probe, the effect of quadratic channels on the squeezing of such probes is very minor. In other words, covariance matrix of displaced thermal states is almost unchanged by a quadratic channel and completely unchanged in the case of passive channels which do not change the mean number of particles in the state. Therefore the first three types of factor play a minor role. Quadratic channels will greatly change the displacement of a displaced thermal state therefore the factor of the last type $\frac{1}{\lambda_k}$ is of great relevance. Higher temperature in displaced thermal states decreases the precision of estimation of quadratic channels. Physically, it is good to have either a hot squeezed state or a cold displaced state as a probe. We illustrate this behavior on the paradigmatic example of phase estimation on Fig.~\ref{fig:thermal_phase} using two one-mode squeezed thermal states and two displaced thermal states.

It is important to point out that every example in the next two sections shows that we fix the mean value of the energy of the probe state; temperature does not account as a resource anymore, and neither does displacement. Optimal probe states will always have its entire energy invested into squeezing. This seems to be a completely general behavior. Nonetheless, we were not able to prove this is always the case for any Gaussian unitary channel. Discussion of this matter can be found in Ref.~\cite{vsafranek2016gaussian}.

\section{Estimation of one-mode Gaussian channels}\label{sec:est_one}
In this section we are going to look at the estimation of one-mode Gaussian unitary channels with purely quadratic generators. For one-mode channels the Hermitian matrix $W$ from Eq.~\eqref{def:Gaussian_unitary} can be naturally parametrized as
\[\label{eq:W1}
W=\begin{bmatrix}
-\theta & i r e^{i \chi} \\
-i r e^{-i \chi} & -\theta
\end{bmatrix}.
\]
For $r=0$ the symplectic matrix $S=e^{iKW}$ represents a one-mode phase-shift $\hat{R}(\theta)=\exp(-i\theta\hat{a}^\dag \hat{a})$, and we will write $S=R(\theta)$. Choosing $\theta=0$ instead, the matrix $S$ represents one-mode squeezing at angle $\chi$, $\hat{S}(r,\chi)=\exp(-\frac{r}{2}(e^{i\chi}\hat{a}^{\dag2}-e^{-i\chi}\hat{a}^{2}))$. Squeezing at angle zero will be denoted as $\hat{S}(r)$ and its symplectic matrix equivalent will be denoted as $S(r)$.

The most general one-mode Gaussian state is the one-mode squeezed rotated displaced thermal state~\cite{Weedbrook2012a}, $\hat{\rho}_0=\hat{D}(\tilde{\bg})\hat{R}(\theta)\hat{S}(r)\hat{\rho}_{\mathrm{th}}\hat{S}^\dag(r)\hat{R}^\dag(\theta)\hat{D}^\dag(\tilde{\bg})$, where $\hat{D}(\tilde{\bg})$ is the Weyl displacement operator defined below Eq.~\eqref{def:Gaussian_unitary}, with the variable of the form $\tilde{\bg}=\norm{{d}}e^{i\phi_d}$. The first and the second moments of this state are
\[\label{eq:general_1mode_state}
\bd_0=(\tilde{\bg},\ov{\tilde{\bg}})^T,\quad \sigma_0=R(\theta)S(r)D_0(\cdots)^\dag,
\]
where $D_0=\mathrm{diag}(\lambda_1,\lambda_1)$. For making expressions shorter we have employed the symbol $(\cdots)$. This symbol represents the same matrices that multiply the diagonal matrix $D_0$ from the left, i.e., in this case $(\cdots)=R(\theta)S(r)$. We will use this general one-mode state as our probe state for one-mode channels, i.e., in Eq.~\eqref{eq:moments_epsilon} we set $S_0=R(\theta)S(r)$.

\subsection{Estimation of a channel combining squeezing and phase change}
First we are going to study a general one-mode Gaussian channel which combines both phase-change and squeezing in an arbitrary direction. Results for the phase-changing and squeezing channel will then be obtained as special cases. We construct this general channel by substituting $\theta\rightarrow\omega_p\epsilon$, $r\rightarrow \omega_s\epsilon$ to Eq.~\eqref{eq:W1}. The resulting symplectic matrix $S_\epsilon:=e^{iKW}$ then represents an encoding operator $\hat{S}_\epsilon=\exp((-i\omega_p\hat{a}^\dag \hat{a}-\frac{\omega_s}{2}(e^{i\chi}\hat{a}^{\dag2}-e^{-i\chi}\hat{a}^{2}))\epsilon)$. $\omega_p$ and $\omega_s$ are the frequencies with which the state is rotated and squeezed respectively. We assume these frequencies and the squeezing angle $\chi$ are known, so $\epsilon$ is the only unknown parameter we are trying to estimate. Using the general probe state~\eqref{eq:general_1mode_state} and methods from Sec.~\ref{sec:gen_framework} we derive the quantum Fisher information,
\begin{widetext}
\[\label{eq:one_mode_channel_QFI}
\begin{split}
H(\epsilon)&=\frac{4\lambda_1^2}{1+\lambda_1^2}\Big(\omega_s^2\big(\cos^2(2\theta+\chi)+\cosh^2(2r)\sin^2(2\theta+\chi)\big)+\omega_p^2\sinh^2(2r)-\omega_s\omega_p\sin(2\theta+\chi)\sinh(4r)\Big)\\
&+\frac{4\norm{{d}}^2}{\lambda_1}\Big(e^{2r}\big(\omega_s\cos(\theta-\phi_d+\chi)-\omega_p\sin(\theta+\phi_d)\big)^2
+e^{-2r}\big(\omega_s\sin(\theta-\phi_d+\chi)+\omega_p\cos(\theta+\phi_d)\big)^2\Big).
\end{split}
\]
\end{widetext}
Assuming all $\omega_s,\omega_p,r$ are positive, this function clearly achieves its maximum when $\sin(2\theta+\chi)=-1$, $\sin(\theta-\phi_d+\chi)=1$, and $\sin(\theta+\phi_d)=-1$. For example, these conditions are fulfilled when $\theta=-\frac{\chi}{2}-\frac{\pi}{4}$, $\phi_d=\frac{\chi}{2}-\frac{\pi}{4}$, which leads to
\[
\begin{split}
H_{\mathrm{max}}(\epsilon)&=\frac{4\lambda_1^2}{1+\lambda_1^2}\big(\omega_s\cosh(2r)+\omega_p\sinh(2r)\big)^2\\
&+\frac{4\norm{{d}}^2}{\lambda_1}e^{2r}\big(\omega_s+\omega_p\big)^2.
\end{split}
\]
This shows that both displacement and squeezing, if properly oriented, enhance the estimation precision. However, to study what strategy is the best when only a fixed amount of energy is available we use the relation for the mean total number of Bosons,
\[\label{eq:mean_number_of_bosons_one}
n=n_{{d}}+n_{\mathrm{th}}+(1+2n_{\mathrm{th}})\sinh^2r,
\]
where $n_{{d}}:=\norm{{d}}^2$ denotes the mean number of Bosons coming from the displacement. Together with the relation $\lambda_1=1+2n_{\mathrm{th}}$ we derive
\[
\begin{split}
&H_{\!\mathrm{max}}\!(\!\epsilon\!)\!\!=\!\!\frac{2\Big(\!\omega_s(2n\!-\!2n_{{d}}\!+\!1)\!+\!2\omega_p\sqrt{\!n\!-\!n_{{d}}\!-\!n_{\mathrm{th}}\!}\sqrt{\!n\!+\!1\!-\!n_{{d}}\!+\!n_{\mathrm{th}}\!}\Big)^{\!2}\!\!\!\!}{1\!+\!2n_{\mathrm{th}}(1\!+\!n_{\mathrm{th}})}\\
&\ +\frac{4n_{{d}}\big(2n\!-\!2n_{{d}}\!+\!1\!+\!2\sqrt{\!n\!-\!n_{{d}}\!-\!n_{\mathrm{th}}\!}\sqrt{\!n\!+\!1\!-\!n_{{d}}\!+\!n_{\mathrm{th}}\!}\big)^2\!\!\!}{(1\!+\!2n_{\mathrm{th}})^2}\!(\omega_s\!+\!\omega_p)^2
\end{split}
\]
Keeping $n$ fixed, the maximum is achieved when $n_{\mathrm{th}}=n_{{d}}=0$, i.e., when all available energy is invested into squeezing, which coincides with some special cases~\cite{Aspachs2008a,Gaiba2009a}. The quantum Fisher information then reaches the Heisenberg limit,
\[
H_{\mathrm{max}}(\epsilon)=2\big(\omega_s(2n+1)+\omega_p2\sqrt{n}\sqrt{1+n}\big)^2.
\]
On the other hand, if we decide to invest only into the displacement (which corresponds to the coherent probe state), i.e., $n=n_{{d}}$, we obtain the shot-noise limit $H_{\mathrm{max}}(\epsilon)=2\omega_s^2+4n(\omega_s+\omega_p)^2$.

\subsection{Estimation of a phase-changing channel}
The quantum Fisher information for the phase-changing channel $\hat{R}(\epsilon)$ is readily obtained from Eq.~\eqref{eq:one_mode_channel_QFI} by setting $\omega_s=0$, $\omega_p=1$,
\[\label{eq:phase_changing_channel}
\begin{split}
H(\epsilon)&=\frac{4\lambda_1^2}{1+\lambda_1^2}\sinh^2(2r)\\
&+\frac{4\norm{{d}}^2}{\lambda_1}\Big(e^{2r}\sin^2(\theta+\phi_d)+e^{-2r}\cos^2(\theta+\phi_d)\Big).
\end{split}
\]
The maximum value is achieved when $\norm{\sin(\theta+\phi_d)}=1$, i.e., for example for $\theta=\pi/2-\phi_d$. This demonstrates that the initial rotation of the squeezed thermal state, or in other words, the angle of squeezing, is irrelevant as long as the displacing is applied in the direction where the squeezed state is stretched. Setting $n_{\mathrm{th}}=n_{{d}}=0$, we obtain the Heisenberg limit $H_{\mathrm{max}}(\epsilon)=8n(n+1)$, which generalizes the precision bound found in~\cite{Aspachs2008a} to any one-mode squeezed Gaussian state, and $n=n_{{d}}$ gives the shot-noise limit $H_{\mathrm{max}}(\epsilon)=4n$. To conclude, the optimal state for phase-estimation is any squeezed thermal state which is displaced in the direction in which it is stretched. The optimal temperature depends on the ratio of the initial squeezing and on the amount of displacing, given by the solution of $\frac{\lambda_1^3}{(\lambda_1^2+1)^2}=\frac{\norm{{d}}^2e^{2r}}{2\sinh^2(2r)}$. When only a finite amount of energy is available for the probe state, the optimal state is any squeezed vacuum. The phase estimation using various probe states is depicted on Fig.~\ref{fig:thermal_phase}.

\subsection{Estimation of a one-mode squeezing channel}\label{subsec:one-mode_squeezing}
The quantum Fisher information for the squeezing channel $\hat{S}(\epsilon,\chi)$ is obtained from Eq.~\eqref{eq:one_mode_channel_QFI} by setting $\omega_s=1$, $\omega_p=0$,
\[\label{eq:one_mode_squeezing_channel}
\begin{split}
H(\epsilon)&=\frac{4\lambda_1^2}{1+\lambda_1^2}\big(\cos^2(2\theta+\chi)+\cosh^2(2r)\sin^2(2\theta+\chi)\big)\\
&+\frac{4\norm{{d}}^2}{\lambda_1}\big(e^{2r}\!\cos^2(\theta\!-\!\phi_d\!+\!\chi)
+e^{-2r}\!\sin^2(\theta\!-\!\phi_d\!+\!\chi)\big).
\end{split}
\]
The maximum is reached when $\norm{\sin(2\theta+\chi)}=1$ and $\norm{\cos(\theta-\phi_d+\chi)}=1$, which occurs for example for $\theta=\pi/4-\chi/2$, $\phi_d=\pi/4+\chi/2$. To achieve the maximal precision we need to rotate the squeezed thermal state by $\pi/4$ from the direction of the squeezing channel we want to estimate, and again as in case of the phase-changing channel, to displace it in the direction in which the squeezed state is stretched. This result generalizes the bounds derived in~\cite{Gaiba2009a} and~\cite{Safranek2015b}, in which the squeezing channels with $\chi=\pi/2$ and $\chi=0$ were studied, respectively. Setting $n_{\mathrm{th}}=n_{{d}}=0$ we obtain the Heisenberg limit $H_{\mathrm{max}}(\epsilon)=2(2n+1)^2$, while $n=n_{{d}}$ gives the shot-noise limit $H_{\mathrm{max}}(\epsilon)=2(2n+1)$. Leading orders of this scaling also correspond to the results from the papers using the global estimation theory~\cite{Milburn1994a,Chiribella2006a}. In conclusion, to optimally estimate the squeezing channel, we prepare the thermal state, squeeze it $\pi/4$ from the direction in which the channel squeezes, and displace in the direction in which it is stretched. The optimal temperature is given by the solution of $\frac{\lambda_1^3}{(\lambda_1^2+1)^2}=\frac{\norm{{d}}^2e^{2r}}{2\cosh^2(2r)}$. When only a finite amount of energy is available, the optimal strategy is to invest it all into squeezing.

\section{Estimation of two-mode Gaussian channels}\label{sec:est_two}
In this section we are going to study the estimation of two-mode Gaussian unitary channels with purely quadratic generators using a wide class of two-mode mixed probe states and the general two-mode pure state. In the analogy with one-mode Gaussian channels, we parametrize the Hermitian matrix $W$ from Eq.~\eqref{def:Gaussian_unitary} for two-mode channels as
\[\label{eq:W2}
W=\begin{bmatrix}
-\theta_1 & -i\theta_{B}e^{i \chi_{B}}  & i r_1 e^{i \chi_1} & i r_{T} e^{i \chi_{T}} \\
i\theta_{B}e^{-i \chi_{B}} & -\theta_2 & i r_{T} e^{i \chi_{T}} & i r_2 e^{i \chi_2} \\
-i r_1 e^{-i \chi_1} & -i r_{T} e^{-i \chi_{T}} & -\theta_1 & i\theta_{B}e^{-i \chi_{B}}\\
-i r_{T} e^{-i \chi_{T}} & -i r_2 e^{-i \chi_2} & -i\theta_{B}e^{i \chi_{B}} & -\theta_2
\end{bmatrix}.
\]
Setting all parameters apart from $\theta_1$ to zero, the matrix $S=e^{iKW}$ represents the one-mode phase-shift $\hat{R}_1(\theta_1)=\exp(-i\theta_1\hat{a}_1^\dag \hat{a}_1)$, and we write $S=R_1(\theta_1)$. Similarly, for $\theta_2$ we have $S=R_2(\theta_2)$. Setting all parameters apart from $\theta_{B}$ and $\chi_{B}$ to zero, we obtain the general mode-mixing channel $\hat{B}(\theta_{B},\chi_{B})=\exp(\theta_{B}(e^{i\chi_{B}}\hat{a}_1^\dag\hat{a}_2-e^{-i\chi_{B}}\hat{a}_2^\dag\hat{a}_1))$, where $\chi_{B}$ represents the angle of mode-mixing. For $\chi_{B}=0$ we obtain the usual beam-splitter with transmissivity $\tau=\cos^2\theta_{B}$, denoted $\hat{B}(\theta_{B})$. Following the same logic, parameters $r_1$ and $r_2$ represent the one-mode squeezing of the first and the second mode as defined in the previous section, denoted $\hat{S}_1(r_1,\chi_1)$, $\hat{S}_2(r_2,\chi_2)$, and parameter $r_{T}$ represents the two-mode squeezing at angle $\chi_{T}$, $\hat{S}_{T}(r_{T},\chi_{T})=\exp(-r_{T}(e^{i\chi_{T}}\hat{a}_1^\dag\hat{a}_2^\dag-e^{-i\chi_{T}}\hat{a}_1\hat{a}_2))$.

We parametrize a general $2\times2$ unitary matrix as
\[
U_1=\begin{bmatrix}
e^{-i \phi_1} & 0 \\
0 & 1
\end{bmatrix}
\begin{bmatrix}
1 & 0 \\
0 & e^{-i \phi_2}
\end{bmatrix}
\begin{bmatrix}
\cos\theta_2 & \sin\theta_2  \\
-\sin\theta_2 & \cos\theta_2
\end{bmatrix}
\begin{bmatrix}
e^{-i \psi_2} & 0 \\
0 & e^{i \psi_2}
\end{bmatrix}.
\]
An equivalent parametrization is
\[
U_2=\begin{bmatrix}
e^{-i \psi_1} & 0 \\
0 & e^{i \psi_1}
\end{bmatrix}
\begin{bmatrix}
\cos\theta_1 & \sin\theta_1  \\
-\sin\theta_1 & \cos\theta_1
\end{bmatrix}
\begin{bmatrix}
e^{-i \phi_3} & 0 \\
0 & 1
\end{bmatrix}
\begin{bmatrix}
1 & 0 \\
0 & e^{-i \phi_4}
\end{bmatrix}.
\]
We insert these matrices into Eq.~\eqref{def:S_decomposition} to obtain the parametrization of a general two-mode Gaussian state in the phase-space formalism. Matrices with phase parameters $e^{-i \phi_3}$ and $e^{-i \phi_4}$ in the parametrization of $U_2$ will vanish because they commute with the diagonal matrix $D_0$ representing the thermal state in the Williamson's decomposition, Eqs.~\eqref{def:Williamson_decomposition} and \eqref{eq:moments_epsilon}. Matrix $U_1$ has its unitary operator equivalent $\hat{U}_1=\hat{R}_1(\phi_1)\hat{R}_2(\phi_2)\hat{B}(\theta_2)\hat{R}_{\mathrm{as}}(\psi_2)$ (see Appendix~\ref{app:list_of_sympl_matrices}), where we define $\hat{R}_{\mathrm{as}}(\psi):=\hat{R}_1(\psi)\hat{R}_2(-\psi)$. Similarly, $\hat{U}_2=\hat{R}_{\mathrm{as}}(\psi_1)\hat{B}(\theta_1)\hat{R}_1(\phi_3)\hat{R}_2(\phi_4)$. This gives a parametrization of a general two-mode Gaussian state in the density matrix formalism,
\[
\begin{split}
\hat{\rho}_0=&\hat{D}(\tilde{\bg})\hat{R}_1(\phi_1)\hat{R}_2(\phi_2)\hat{B}(\theta_2)\hat{R}_{\mathrm{as}}(\psi_2)\hat{S}_1(r_1)\hat{S}_2(r_2)\\
&\hat{R}_{\mathrm{as}}(\psi_1)\hat{B}(\theta_1)\hat{\rho}_{\mathrm{th}}(\cdots)^\dag,
\end{split}
\]
where the variable in the Weyl displacement operator $\hat{D}(\tilde{\bg})$ is of the form $\tilde{\bg}=(\norm{{d}_1}e^{i\phi_{d1}},\norm{{d}_2}e^{i\phi_{d2}})$. $(\cdots):=\hat{D}(\tilde{\bg})\dots\hat{B}(\theta_1)$ has the same meaning as in Eq.~\eqref{eq:general_1mode_state}. It is possible to find a parametrization of a general three-mode Gaussian state using the same technique~\cite{vsafranek2016gaussian}. An equivalent parametrization of a two-mode Gaussian state can be found in~\cite{serafini2003symplectic}, but we decided to use the above because it requires fewer active transformations (i.e., two squeezing transformations as compared to three), and it ultimately leads to simpler results.

Although analysis with the general two-mode state can be made, the results seem to be too complicated to be used effectively. Also, as the first three operations applied on the thermal state only swap and entangle the symplectic eigenvalues, we do not expect much generality will be lost when not considering them. Moreover, in the case of the isothermal states (which also covers all pure states), such operations do not have any effect. This is why we restrict ourselves to probe states which we write in the covariance matrix formalism as
\begin{subequations}\label{eq:simplified_probe_state}
\begin{gather}
\bd_0=(\tilde{\bg},\ov{\tilde{\bg}})^T,\\
\sigma_0=R_1(\phi_1)R_2(\phi_2)B(\theta)R_{\mathrm{as}}(\psi)S_1(r_1)S_2(r_2)D_0(\cdots)^\dag,
\end{gather}
\end{subequations}
where $D_0=\mathrm{diag}(\lambda_1,\lambda_2,\lambda_1,\lambda_2)$. Also, since using mixed states cannot improve the
quality of estimation when fixing the energy of the probe state, the optimal states are always pure. As Eq.~\eqref{eq:simplified_probe_state} encompasses all pure states, it is enough to use this restricted class of states to find the optimal.

\subsection{Estimation of two-mode squeezing channels}
First we are going to study the optimal states for the estimation of the two-mode squeezing channel $\hat{S}_{T}(\epsilon,\chi)$, assuming the direction of squeezing $\chi$ is known. Using the state from Eq.~\eqref{eq:simplified_probe_state} we find only two cases which lead to significantly different results. In the first case a beam-splitter is not used ($\theta=0$) in the preparation process, which corresponds to using two simultaneously sent, but non-entangled single-mode squeezed probe states. In the second case the balanced beam-splitter is used ($\theta=\pi/4$), which corresponds to using two-mode squeezed-type probe states. The full expression for the quantum Fisher information is a mixture of these two qualitatively different cases and can be found in Appendix~\ref{app:full_expressions}.

\subsubsection{Two-mode squeezing channel: Using two nonentangled single-mode squeezed Gaussian states}
Assuming $\theta=0$ in the probe state~\eqref{eq:simplified_probe_state}, without loss of generality we can also set $\psi=0$. The resulting quantum Fisher information for the estimation of a two-mode squeezing channel reads
\begin{widetext}
\[\label{eq:QFI_nonentangled_twomode}
\begin{split}
H(\epsilon)&\!=\!\frac{2(\lambda_1+\lambda_2)^2}{\lambda_1\lambda_2+1}\big(\!\cos^2\!\!\phi_\chi\cosh^2(\!r_1\!-\!r_2\!) +\sin^2\!\!\phi_\chi\cosh^2(\!r_1\!+\!r_2\!)\big)
+\frac{2(\lambda_1-\lambda_2)^2}{\lambda_1\lambda_2-1}\big(\!\cos^2\!\!\phi_\chi\sinh^2(\!r_1\!-\!r_2\!) +\sin^2\!\!\phi_\chi\sinh^2(\!r_1\!+\!r_2\!)\big)\\
&+\frac{4\norm{{d}_2}^2}{\lambda_1}\big(e^{2r_1}\cos^2\!\!\phi_{1\chi}+e^{-2r_1}\sin^2\!\!\phi_{1\chi}\big)
+\frac{4\norm{{d}_1}^2}{\lambda_2}\big(e^{2r_2}\cos^2\!\!\phi_{2\chi}+e^{-2r_2}\sin^2\!\!\phi_{2\chi}\big),
\end{split}
\]\end{widetext}
where we have denoted $\phi_\chi:=\phi_1+\phi_2+\chi$, $\phi_{1\chi}:=\phi_1-\phi_{d2}+\chi$, $\phi_{2\chi}:=\phi_2-\phi_{d1}+\chi$. The presence of mixed temperature terms in the expression shows that using non-entangled squeezed Gaussian states yields the possibility of the temperature-enhanced estimation. Assuming both $r_1$ and $r_2$ are positive, the maximum is reached for $\phi_\chi=\frac{\pi}{2}$ and $\phi_{1\chi}=\phi_{2\chi}=0$, which leads to
\[\label{eq:max_QFI_nonentangled_twomode}
\begin{split}
H_{\!\mathrm{max}}\!(\epsilon)\!\!&=\!\!\frac{2(\lambda_1\!\!+\!\!\lambda_2)^2\!\!}{\lambda_1\lambda_2\!+\!1}\cosh^2\!(\!r_1\!+\!r_2\!)
+\frac{2(\lambda_1\!\!-\!\!\lambda_2)^2\!\!}{\lambda_1\lambda_2\!-\!1}\sinh^2\!(\!r_1\!+\!r_2\!)\\
&+\frac{4\norm{{d}_2}^2}{\lambda_1}e^{2r_1}
+\frac{4\norm{{d}_1}^2}{\lambda_2}e^{2r_2}.
\end{split}
\]
These conditions are fulfilled, for example, for $\phi_1=\phi_2=\frac{\pi}{4}-\frac{\chi}{2}$, $\phi_{d1}=\phi_{d2}=\frac{\pi}{4}+\frac{\chi}{2}$, which is in complete analogy with the optimal states for the one-mode squeezing channel from Sec.~\ref{subsec:one-mode_squeezing}. This means that we can effectively probe the two-mode squeezing channel by two simultaneously sent copies of the optimal states for the one-mode squeezing channel. Note the mixed term $\frac{4\norm{{d}_2}^2}{\lambda_1}e^{2r_1}$, which combines the squeezing of one mode and enhances it by the displacement of the other mode, demonstrating the entangling nature of the two-mode squeezing channel.

To study the optimal states when only a finite amount of energy is available, we use the two-mode equivalent of Eq.~\eqref{eq:mean_number_of_bosons_one} for the mean total number of Bosons,
\[\label{eq:mean_number_of_bosons_two}
n=n_{{d}_1}+n_{\mathrm{th}1}+\lambda_1\sinh^2r_1
+n_{{d}_2}+n_{\mathrm{th}2}+\lambda_2\sinh^2r_2,
\]
where $n_{{d}_i}:=\norm{{d}_i}^2$, and $\lambda_i=1+2n_{\mathrm{th}i}$, $i=1,2$. Maximizing the quantum Fisher information while keeping the $n$ fixed we find that the maximum is achieved when the initial squeezings are equal, $r_1=r_2$, and all energy is invested into squeezing, 
reaching the Heisenberg limit $H_{\mathrm{max}}(\epsilon)=4(n+1)^2$. If we invest only into the displacement, $n=n_{{d}_1}+n_{{d}_2}$, independently of the ratio ${n_{{d}_1}}/{n_{{d}_2}}$ we obtain the shot-noise limit $H_{\mathrm{max}}(\epsilon)=4(n+1)$.

\subsubsection{Two-mode squeezing channel: Using beam splitter in the preparation process}
Setting $\theta=\frac{\pi}{4}$ in probe state~\eqref{eq:simplified_probe_state} and using the same notation as in Eq.~\eqref{eq:QFI_nonentangled_twomode}, we derive the quantum Fisher information for the estimation of the two-mode squeezing channel
\begin{widetext}
\[
\begin{split}
H(\epsilon)&=\frac{4\lambda_1^2}{\lambda_1^2+1}\big(\cos^2(\phi_\chi+2\psi)+\sin^2(\phi_\chi+2\psi)\cosh(2r_1)\big)
+\frac{4\lambda_2^2}{\lambda_2^2+1}\big(\cos^2(\phi_\chi-2\psi)+\sin^2(\phi_\chi-2\psi)\cosh(2r_2)\big)\\
&+\frac{2}{\lambda_1}\Big(e^{2r_1}\big(\norm{{d}_1}\cos(\phi_{2\chi}+\psi)-\norm{{d}_2}\cos(\phi_{1\chi}+\psi)\big)^2
+e^{-2r_1}\big(\norm{{d}_1}\sin(\phi_{2\chi}+\psi)-\norm{{d}_2}\sin(\phi_{1\chi}+\psi)\big)^2\Big)\\
&+\frac{2}{\lambda_2}\Big(e^{2r_2}\big(\norm{{d}_1}\cos(\phi_{2\chi}-\psi)+\norm{{d}_2}\cos(\phi_{1\chi}-\psi)\big)^2
+e^{-2r_2}\big(\norm{{d}_1}\sin(\phi_{2\chi}-\psi)+\norm{{d}_2}\sin(\phi_{1\chi}-\psi)\big)^2\Big).\\
\end{split}
\]
\end{widetext}
Unlike the previous case given by Eq.~\eqref{eq:QFI_nonentangled_twomode}, the lack of mixed temperature terms in this expression shows that using a beam splitter in the preparation process prohibits the temperature-enhanced estimation. Moreover, the maximum can no longer be identified easily. For example, when both $r_1$, $r_2$ are positive and $\frac{e^{2r_2}}{\lambda_2}\geq\frac{e^{2r_1}}{\lambda_1}$, one of the optimal states is given by $\phi_\chi=\frac{\pi}{2}$, $\psi=\phi_{1\chi}=\phi_{2\chi}=0$ and leads to the quantum Fisher information
\[\label{eq:max_QFI_entangled_twomode}
\begin{split}
H_{\mathrm{max}}(\epsilon)&=\frac{4\lambda_1^2}{\lambda_1^2+1}\cosh^2(2r_1)
+\frac{4\lambda_2^2}{\lambda_2^2+1}\cosh^2(2r_2)\\
&+\frac{2}{\lambda_1}(\norm{{d}_1}-\norm{{d}_2})^2e^{2r_1}
+\frac{2}{\lambda_2}(\norm{{d}_1}+\norm{{d}_2})^2e^{2r_2}.
\end{split}
\]
For $\lambda_1=\lambda_2=1$ and $r_1=r_2$ such an optimal state reduces to the two single-mode squeezed states. In contrast, for $r_1\leq0$, $r_2\geq0$, and $\frac{e^{2r_2}}{\lambda_2}\geq\frac{e^{-2r_1}}{\lambda_1}$, the optimal state is given by
$\phi_\chi=0$, $\psi=\phi_{1\chi}=\phi_{2\chi}=\frac{\pi}{4}$ and leads to
\[\label{eq:max_QFI_entangled_twomodeb}
\begin{split}
H_{\mathrm{max}}(\epsilon)&=\frac{4\lambda_1^2}{\lambda_1^2+1}\cosh^2(2r_1)
+\frac{4\lambda_2^2}{\lambda_2^2+1}\cosh^2(2r_2)\\
&+\frac{2}{\lambda_1}(\norm{{d}_1}-\norm{{d}_2})^2e^{-2r_1}
+\frac{2}{\lambda_2}(\norm{{d}_1}+\norm{{d}_2})^2e^{2r_2}.
\end{split}
\]
For $\lambda_1=\lambda_2=1$ and $r_1=r_2$ such optimal state reduces to the two-mode squeezed probe state.

The difference between formulas~\eqref{eq:max_QFI_entangled_twomode} and~\eqref{eq:max_QFI_entangled_twomodeb} is only in the use of the displacement. Non-displaced probe states reach the same precision independently of the sign of the squeezing parameters. Maximizing the quantum Fisher information for a fixed amount of energy in the probe state we arrive at the very same conclusions as in the case of nonentangled states, i.e., the optimal state is obtained when all energy is invested into squeezing and squeezing parameters are equal, giving the same Heisenberg limit. Here, however, the last part of the expression, $(\norm{{d}_1}+\norm{{d}_2})^2e^{2r_2}$, combines the displacements by the mixed term $2\norm{{d}_1}\norm{{d}_2}$ which is not present when using non-entangled probe states. This may be useful if, for some reason, we want to squeeze only one of the two modes (for example when the apparatus for creating squeezed states is expensive or difficult to build). Sending a coherent state in the other mode then enhances the estimation in a non-linear way.

Now, let us see whether it is more effective to use an entangled state or two one-mode squeezed states as a probe for the estimation of the two-mode squeezing channel. To do that we compare the precision of estimation where beam-splitter has and has not been used in the preparation process. Assuming both modes are pure, $\lambda_1=\lambda_2=1$, and subtracting Eq.~\eqref{eq:max_QFI_nonentangled_twomode} from Eq.~\eqref{eq:max_QFI_entangled_twomode}, we obtain 
\[\label{eq:difference_in_r1_r2}
\begin{split}
&H_{\!\!\mathrm{max}BS}\!(\epsilon)-H_{\!\!\mathrm{max}\cancel{BS}}\!(\epsilon)=4\cosh(2(r_1\!+\!r_2))\sinh^2(r_2\!-\!r_1)\\
&\ \ \ +4(\norm{{d}_2}^2+2\norm{{d}_1}\norm{{d}_2}-\norm{{d}_1}^2)e^{r_1+r_2}\sinh(r_2-r_1).
\end{split}
\]
This shows that unless the displacement of the first mode $\norm{{d}_1}$ is very large, using a beam-splitter exploits the difference in squeezing parameters more effectively. The advantage however vanishes when the optimal strategy ($r_1=r_2$) is used.

\subsubsection{Two-mode squeezing channel: Using one-mode Gaussian states}
In the previous sections we considered two-mode Gaussian probe states for the estimation of two-mode channels. But is probing them with the two-mode states really necessary? What precision could be achieved by using only the one-mode state as a probe? Mathematically, we represent such one-mode Gaussian probes by a two-mode Gaussian state where the first mode is the most general single-mode state and the second mode is vacuum, $\hat{\rho}_0=\hat{D}(\norm{{d}_1}e^{i\phi_{d1}})\hat{R}(\phi_1)\hat{S}(r_1)\hat{\rho}_{\mathrm{th}}(\cdots)^\dag\otimes\ket{0}\bra{0}$. The quantum Fisher information is easily obtained from Eq.~\eqref{eq:QFI_nonentangled_twomode} by setting $r_2=d_2=0$, $\lambda_2=1$,
\[
H(\epsilon)=2\lambda_1\cosh(2r_1)+4\norm{{d}_1}^2+2,
\]
which gives the shot-noise limit $H(\epsilon)=4(n+1)$ independently of how energy is distributed among squeezing, displacement, and temperature. Although it is possible to use one-mode states to estimate the two-mode squeezing channel, it is not effective.

\subsection{Estimation of mode-mixing channels}
In this section we study optimal states for the estimation of the mode-mixing channel $\hat{B}(\epsilon,\chi)$, assuming the `direction' of mixing $\chi$ is known, again with the probe state given by Eq.~\eqref{eq:simplified_probe_state}. Similarly to the previous section, we show the case when a beam-splitter has been used in the preparation process, and when non-entangled states have been used instead. In contrast to the two-mode squeezing channel, where the optimal probe state always depended on the squeezing angle $\chi$, here we identify a universal probe state which achieves the optimal scaling for any mode-mixing angle $\chi$. The full expression for the quantum Fisher information for the estimation of mode-mixing channels can be found in Appendix~\ref{app:full_expressions}.

\subsubsection{Mode-mixing channel: Using two nonentangled single-mode squeezed Gaussian states}
Assuming $\theta=0$ in Eq.~\eqref{eq:simplified_probe_state}, and without loss of generality also $\psi=0$, we derive the quantum Fisher information for the estimation of the mode-mixing channel,
\begin{widetext}
\[\label{eq:QFI_nonentangled_bs}
\begin{split}
H(\epsilon)&\!=\!\frac{2(\lambda_1+\lambda_2)^2}{\lambda_1\lambda_2+1}\big(\!\cos^2\!\!\phi_\chi\sinh^2(\!r_1\!-\!r_2\!) +\sin^2\!\!\phi_\chi\sinh^2(\!r_1\!+\!r_2\!)\big)
+\frac{2(\lambda_1-\lambda_2)^2}{\lambda_1\lambda_2-1}\big(\!\cos^2\!\!\phi_\chi\cosh^2(\!r_1\!-\!r_2\!) +\sin^2\!\!\phi_\chi\cosh^2(\!r_1\!+\!r_2\!)\big)\\
&+\frac{4\norm{{d}_2}^2}{\lambda_1}\big(e^{2r_1}\cos^2\!\!\phi_{1\chi}+e^{-2r_1}\sin^2\!\!\phi_{1\chi}\big)
+\frac{4\norm{{d}_1}^2}{\lambda_2}\big(e^{2r_2}\cos^2\!\!\phi_{2\chi}+e^{-2r_2}\sin^2\!\!\phi_{2\chi}\big),
\end{split}
\]\end{widetext}
where we have denoted $\phi_\chi:=\phi_1-\phi_2+\chi$, $\phi_{1\chi}:=\phi_1+\phi_{d2}+\chi$, $\phi_{2\chi}:=\phi_2+\phi_{d1}-\chi$. Note that the difference between this formula and Eq.~\eqref{eq:QFI_nonentangled_twomode} lies only in the different definitions of  $\phi_\chi$, $\phi_{1\chi}$, $\phi_{2\chi}$, and swapping $\cosh\leftrightarrow\sinh$. Again, temperature-enhanced estimation is possible. For positive $r_1$ and $r_2$ the maximum is reached when $\phi_\chi=\frac{\pi}{2}$ and $\phi_{1\chi}=\phi_{2\chi}=0$,
\[\label{eq:max_QFI_nonentangled_bs}
\begin{split}
H_{\mathrm{max}}(\epsilon)\!&=\!\frac{2(\lambda_1\!\!+\!\!\lambda_2)^2\!\!}{\lambda_1\lambda_2\!+\!1}\sinh^2(\!r_1\!+\!r_2\!)
+\frac{2(\lambda_1\!\!-\!\!\lambda_2)^2\!\!}{\lambda_1\lambda_2\!-\!1}\cosh^2(\!r_1\!+\!r_2\!)\\
&+\frac{4\norm{{d}_2}^2}{\lambda_1}e^{2r_1}
+\frac{4\norm{{d}_1}^2}{\lambda_2}e^{2r_2}.
\end{split}
\]
These conditions are fulfilled for $\phi_1=\frac{\pi}{4}-\frac{\chi}{2}$, $\phi_{d1}=\frac{\pi}{4}+\frac{\chi}{2}$, $\phi_2=-\frac{\pi}{4}+\frac{\chi}{2}$, $\phi_{d2}=-\frac{\pi}{4}-\frac{\chi}{2}$. Using Eq.~\eqref{eq:mean_number_of_bosons_two} we show that the energy-optimal probe state is obtained, again as in the estimation of the two-mode squeezing channel, when the entire energy is uniformly distributed among squeezing parameters, reaching the Heisenberg limit $H_{\mathrm{max}}(\epsilon)=4n(n+2)$. Investing only in the displacement, $n=n_{{d}_1}+n_{{d}_2}$, we obtain the shot-noise limit $H_{\mathrm{max}}(\epsilon)=4n$.

\subsubsection{Mode-mixing channel: Using beam splitter in the preparation process}
Setting $\theta=\frac{\pi}{4}$ in Eq.~\eqref{eq:simplified_probe_state} we derive the quantum Fisher information
\begin{widetext}
\[\label{eq:QFI_entangled_BS}
\begin{split}
&H(\epsilon)=4\sin^2\!\!\phi_\chi\bigg(\frac{\lambda_1^2}{\lambda_1^2+1}\sinh^2(2r_1)
+\frac{\lambda_2^2}{\lambda_2^2+1}\sinh^2(2r_2)\bigg)\\
&\!\!+\!2\!\cos^2\!\!\phi_\chi\bigg(\!\!\frac{(\lambda_1\!+\!\lambda_2)^2}{\lambda_1\lambda_2\!+\!1}\!\big(\!\!\cos^2\!(\!2\psi\!)\!\sinh^2(\!r_1\!-\!r_2\!)\!+\!\sin^2\!(\!2\psi\!)\!\sinh^2(\!r_1\!+\!r_2\!)\!\big)
\!+\!\frac{(\lambda_1\!-\!\lambda_2)^2}{\lambda_1\lambda_2\!-\!1}\!\big(\!\!\cos^2\!(\!2\psi\!)\!\cosh^2(\!r_1\!-\!r_2\!)\!+\!\sin^2\!(\!2\psi\!)\!\cosh^2(\!r_1\!+\!r_2\!)\!\big)\!\!\bigg)\\
&\!\!+\frac{2}{\lambda_1}\Big(e^{2r_1}\big(\norm{{d}_1}\cos(\phi_{2\chi}+\psi)+\norm{{d}_2}\cos(\phi_{1\chi}+\psi)\big)^2
+e^{-2r_1}\big(\norm{{d}_1}\sin(\phi_{2\chi}+\psi)+\norm{{d}_2}\sin(\phi_{1\chi}+\psi)\big)^2\Big)\\
&\!\!+\frac{2}{\lambda_2}\Big(e^{2r_2}\big(\norm{{d}_1}\cos(\phi_{2\chi}-\psi)-\norm{{d}_2}\cos(\phi_{1\chi}-\psi)\big)^2
+e^{-2r_2}\big(\norm{{d}_1}\sin(\phi_{2\chi}-\psi)-\norm{{d}_2}\sin(\phi_{1\chi}-\psi)\big)^2\Big),\\
\end{split}
\]\end{widetext}
where we use the same notation as in Eq.~\eqref{eq:QFI_nonentangled_bs}. In contrast to the estimation of the two-mode squeezing channel, in the estimating of mode-mixing channels, the use of a beam-splitter in the preparation process does not prevent us from exploiting the temperature-enhanced estimation, which can be done by choosing $\phi_\chi=0$. Choosing $\phi_\chi=\frac{\pi}{2}$ leads to the case where the temperature difference cannot be used, but in the analogy of Eq.~\eqref{eq:difference_in_r1_r2} the difference $r_2-r_1$ is used more effectively. For both these strategies optimizing for the fixed amount of energy of the initial state leads to the same conclusions and the same scaling with the total number of particles as in the case of the two non-entangled probe states.

\subsubsection{Mode-mixing channel: Pure states and the universal state}
For mode-mixing channels we find a unique phenomenon which does not occur with the squeezing channels, and which can be exploited only when using a beam-splitter in the preparation process. Setting $\lambda_1=\lambda_2=1$, $r_1=r_2=r$, $\psi=\frac{\pi}{4}$, and $\phi_1+\phi_2+\phi_{d1}+\phi_{d2}=-\frac{\pi}{2}$ in Eq.~\eqref{eq:QFI_entangled_BS}, we derive
\[
\begin{split}
H(\epsilon)&=4\sinh^2(2r)\\
&+4\Big(\big(\norm{{d}_1}^2+\norm{{d}_2}^2\big)\cosh(2r)+2\norm{{d}_1}\norm{{d}_2}\sinh(2r)\Big).
\end{split}
\]
Any free parameter has, at this point, not been set to be dependent on the `direction' of the mode-mixing $\chi$. Also, the leading order here is identical to the energy-optimal probe states. In other words, we have found an optimal and universal probe state for the mode-mixing channels $\hat{B}(\epsilon,\chi)$. If we set the initial displacement $\bd_0$ to zero, according to Eq.~\eqref{eq:simplified_probe_state} this probe state becomes the two-mode squeezed vacuum in the direction of $\chi_T=\frac{\pi}{2}$, $\hat{\rho}_{0}
=\hat{S}_{T}(r,\tfrac{\pi}{2})\ket{0}\ket{0}\bra{0}\bra{0}\hat{S}_{T}^\dag(r,\tfrac{\pi}{2})$.

\subsubsection{Mode-mixing channel: Using one-mode Gaussian states}

\begin{figure}[t!]
\centering
\includegraphics[width=1\linewidth]{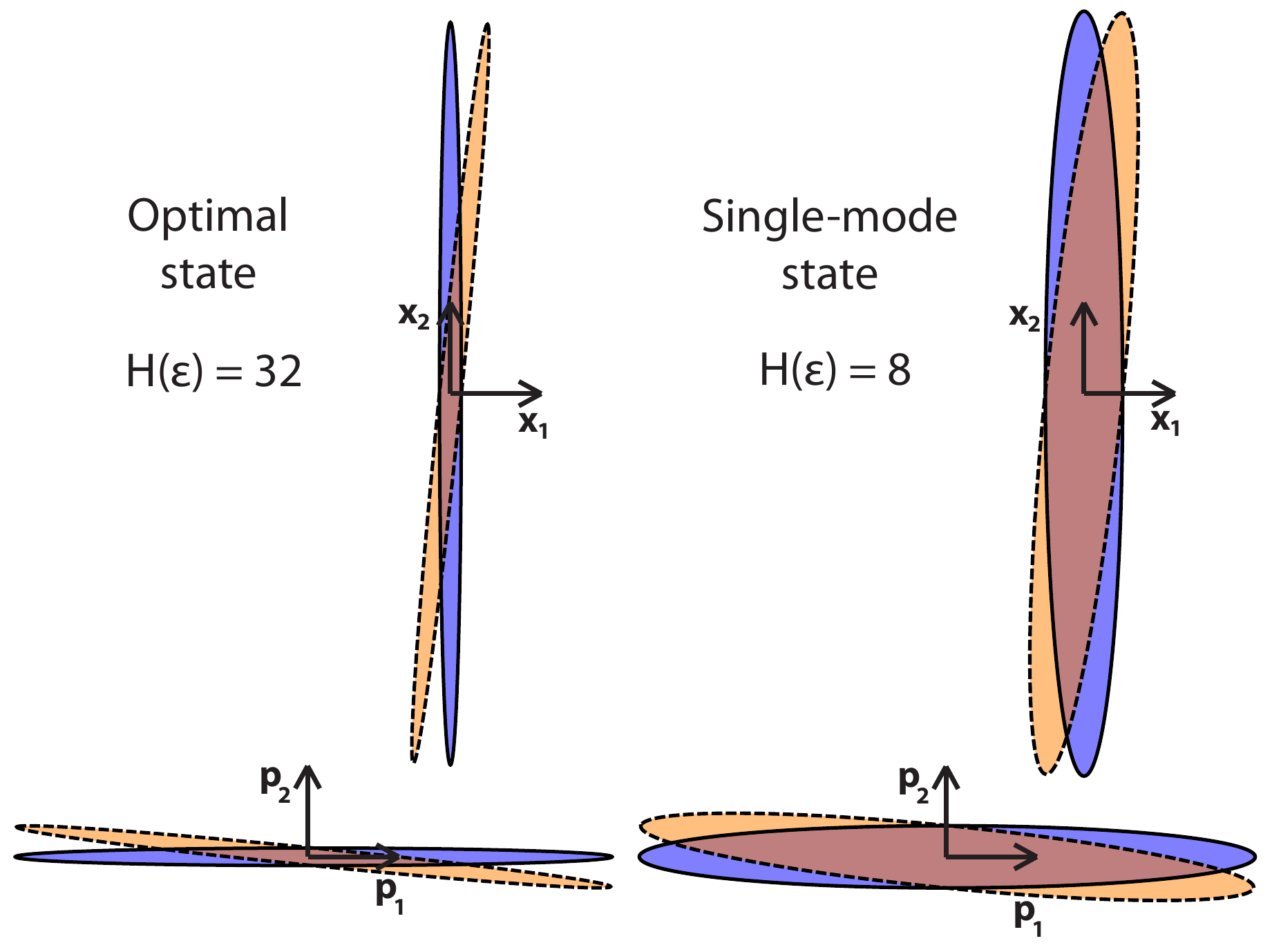}
\caption{Estimation of the beam-splitter $\hat{B}(\epsilon)$ around point $\epsilon=0$ using one of the optimal states, 
$\hat{\rho}=\hat{S}(r)\ket{0}\bra{0}\hat{S}^\dag(r)\otimes\hat{S}(-r)\ket{0}\bra{0}\hat{S}^\dag(-r)$ representing two one-mode squeezed states that are squeezed in orthogonal directions, and the one-mode squeezed state $\hat{\rho}=\hat{S}(r_1)\pro{0}{0}\hat{S}^\dag(r_1)\otimes\ket{0}\bra{0}$, both with the same mean energy $n=2$. We plot the real form marginal covariance matrices showing correlations between positions in the first and the second mode $x_1$ and $x_2$, and momenta $p_1$ and $p_2$ in the real form phase-space, before(blue with full line) and after(orange with dashed line) beam-splitter $\hat{B}(0.1)$ has been applied. There are no correlations between position and momentum. Clearly, the optimal state is more sensitive to the channel allowing for a better estimation of the parameter~$\epsilon$.}\label{fig:one_and_optimal}
\end{figure}

The quantum Fisher information for the estimation of the mode-mixing channel using the most general single-mode state $\hat{\rho}_0=\hat{D}(\norm{{d}_1}e^{i\phi_{d1}})\hat{R}(\phi_1)\hat{S}(r_1)\hat{\rho}_{\mathrm{th}}(\cdots)^\dag\otimes\ket{0}\bra{0}$ is obtained by setting $r_2=d_2=0$, $\lambda_2=1$ in Eq.~\eqref{eq:QFI_nonentangled_bs},
\[
H(\epsilon)=2\lambda_1\cosh(2r_1)+4\norm{{d}_1}^2-2,
\]
which always leads to the shot-noise limit $H(\epsilon)=4n$. For an illustration of how a one-mode state compares to the optimal state, see Fig.~\ref{fig:one_and_optimal}.

\section{Role of entanglement and the Heisenberg limit}
In this section we first show why it is usually thought that entanglement in the probe state is necessary to achieve the Heisenberg limit, and why this reasoning is not applicable in the continuous variable states known as Gaussian states. There are numerous possible definitions of the Heisenberg limit in quantum metrology~\cite{Giovannetti2011a}. In the presence of an infinite-dimensional
system, the definition of Heisenberg limit is somehow difficult as one
needs to resort to the mean energy of the probe state, instead of using the definition based on the number of
“qubits” employed. However, using such a definition is reasonable for Gaussian states. As we have shown in the previous section, a higher mean value of energy usually leads to a higher precision in estimation and thus can be counted as a resource. For example, in the advanced gravitational wave detector LIGO~\cite{aasi2015advanced} the laser light is recycled which greatly boosts the power of the beam and consequently leads to a higher resolution of the detector.

Sequence of states $\hat{\rho}_m$ with ever-increasing mean value of energy $\lim_{m\rightarrow\infty}\mean{\hat{E}}_{\hat{\rho}_m}=\infty$ is said to reach the Heisenberg limit if and only if there exists a number $c>0$ such that
\[
\lim_{m\rightarrow\infty}\frac{H(\hat{\rho}_m)}{(\mean{\hat{E}}_{\hat{\rho}_m})^2}=c.
\]
In the case of a Bosonic system the operator $\hat{E}$ which measures the energy of the probe state is up to a scaling constant identical to the total number operator, $\hat{E}\equiv\hat{N}$.

We first consider a Hilbert space $\mathcal{H}$ such that for every state $\hat{\rho}\in\mathcal{H}$ the quantum Fisher information is bounded by the same value $B_H$, i.e.,
\[\label{eq:bound_H}
\exists B_H>0,\ \ \forall \hat{\rho}\in \mathcal{H},\ \ H(\hat{\rho})\leq B_H.
\]
It is not possible to create a sequence of states from such Hilbert space to achieve the Heisenberg limit, because by definition $\lim_{m\rightarrow\infty}\frac{H(\hat{\rho}_m)}{(\mean{\hat{E}}_{\hat{\rho}_m})^2}\leq\lim_{m\rightarrow\infty}\frac{B_H}{\mean{\hat{E}}_{\hat{\rho}_m}}=0$.
However, we can increase the quantum Fisher information by adding more particles, which corresponds to expanding the Hilbert space. We consider a (fully) separable state
\[\label{eq:construction_separable_states}
\hat{\rho}_m=\sum_ip_i\hat{\rho}_i^{(1)}\otimes\hat{\rho}_i^{(2)}\otimes\cdots\otimes\hat{\rho}_i^{(m)}\in\mathcal{H}^{\otimes m},
\]
where $\sum_ip_i=1$. Assuming that energy of each added state does not go below certain value, i.e.,
\[\label{eq:bound_E}
\exists B_E>0,\ \ \forall i,\ \forall k,\ \ \mean{\hat{E}}_{\hat{\rho}_i^{(k)}}\geq B_E,
\]
and using convexity of the quantum Fisher information and additivity under tensoring~\cite{Toth2014a}, we derive
\[\label{eq:proof_separable_states_QFI}
\lim_{m\rightarrow\infty}\!\frac{H(\hat{\rho}_m)}{(\mean{\hat{E}}_{\hat{\rho}_m})^2}\leq\!\lim_{m\rightarrow\infty}\!\frac{\sum_{i,k}p_iH(\hat{\rho}_i^{(k)})}{(\sum_{i,k}p_i\mean{\hat{E}}_{\hat{\rho}_i^{(k)}}\!)^2}
\leq\!\lim_{m\rightarrow\infty}\!\frac{mB_H}{m^2B_E}=0.
\]
This illustrates that under conditions~\eqref{eq:bound_H} and~\eqref{eq:bound_E}, the construction~\eqref{eq:construction_separable_states} using separable states cannot lead to the Heisenberg limit and entangled states are necessary. This follows the proofs from~\cite{Pezze2009a,Demkowicz2012a,Toth2014a} showing that existence of entanglement in an $m$-qubit state is necessary condition for the scaling of the quantum Fisher information larger than the shot-noise limit.

Although one-qubit Hilbert space, from which the $m$-qubit Hilbert space is created, satisfies Eq.~\eqref{eq:bound_H}, such a condition is no longer satisfied by the Fock space representing a Bosonic system. There are states in the Fock space, such as squeezed states and coherent states, which can lead to an arbitrarily large precision in the estimation. Therefore proof~\eqref{eq:proof_separable_states_QFI} does not apply anymore and entanglement is not necessary. As shown in previous sections, separable states such as squeezed states can also achieve the Heisenberg limit.

In comparison to $m$-qubit systems, which use entanglement as a resource, the resources in Bosonic systems are rather highly superposed states spanning across all infinite-dimensional Hilbert space, while entanglement does not play a significant role anymore.

\section{Concluding remarks and discussion}

In this paper we have exploited recent developments in the theory of metrology and translated the problem of optimal estimation into the more convenient phase-space formalism. This allowed us to systematically study wide classes of Gaussian states for the estimation of Gaussian unitary channels. Using this approach we managed to find optimal states for the most common channels.

We found that for every channel we studied the optimal states are either squeezed or two-mode squeezed states. Further, the entanglement of the probe state does not play any significant role, which corresponds to the findings of~\cite{Gaiba2009a,Friis2015a}. This is not in contradiction with some previous studies showing that entanglement is necessary to achieve the Heisenberg limit~\cite{Pezze2009a,Demkowicz2012a}, because assumptions taken there do not apply anymore to the Fock space describing a Bosonic system.

In estimating parameters of phase-changing, one-mode squeezing, mode-mixing, and two-mode squeezing channels ($\hat{R},\hat{S},\hat{B},\hat{S}_T$ respectively), the quantum Fisher information reaches the Heisenberg limits
\begin{subequations}
\begin{align}
H_{R}(\epsilon)&=2\sinh^2(2r)=8n(n+1),\\
H_{S}(\epsilon)&=2\cosh^2(2r)=2(2n+1)^2,\\
H_{B}(\epsilon)&=4\sinh^2(2r)=4n(n+2),\\
H_{S_T}(\epsilon)&=4\cosh^2(2r)=4(n+1)^2,
\end{align}
\end{subequations}
where $r$ denotes the squeezing of one of the modes in the probe state, and $n$ is the mean total number of particles of the probe state.
These results generalize the precision bounds found in~\cite{Milburn1994a,Chiribella2006a,Monras2006a,Aspachs2008a,Gaiba2009a}.

Alternatively, if we choose coherent states as probe states, we obtain the shot-noise limits
\begin{subequations}
\begin{align}
H_{R}(\epsilon)&=4n,\\
H_{S}(\epsilon)&=2(2n+1),\\
H_{B}(\epsilon)&=4n,\\
H_{S_T}(\epsilon)&=4(n+1).
\end{align}
\end{subequations}
These are the same limits we find when using any one-mode state to probe two-mode Gaussian channels.

Authors of~\cite{Aspachs2008a} showed that temperature of the probe state may enhance the estimation precision by a factor of 2, and authors of~\cite{Gaiba2009a} explored how temperature acts in the estimation of mode-mixing channels. We demonstrated that effects of temperature are generic. Independent of which Gaussian unitary channel is probed, the effects of temperature always come in multiplicative factors of four types. The first three appear when the channel changes the squeezing or the orientation of squeezing of the probe state. The first one accounts for the absolute number of thermal Bosons in each mode and corresponds to the one found in~\cite{Aspachs2008a}. Two of them take into account differences between thermal Bosons in each mode. Larger differences then lead to higher precision in the estimation, while the enhancement factor scales with the ratio of the number of thermal Bosons $\frac{n_{{\mathrm{th}}i}}{n_{{\mathrm{th}}j}}$, for $n_{{\mathrm{th}}i}\gg n_{{\mathrm{th}}j}\gg 0$. The last factor is of the form $(2n_{{\mathrm{th}}i}+1)^{-1}$ and it appears when the Gaussian channel changes the displacement of the probe state.

The main goal of this paper was to show how different aspects of the probe states affect the estimation, and to provide a framework that can be effectively used to study optimal probe states for the construction of new-era quantum detectors. In addition to applications for existing gravitational wave detectors~\cite{Abbott2004a,Caron1995a}, our results may be useful for designing new gravimeters~\cite{Snadden1998a,Altin2013a,Sabin2014a}, climate probes~\cite{Tapley2004a}, or for the estimation of space-time parameters~\cite{Danzmann1996a,Everitt2011a,Bruschi2014a}.


\emph{Acknowledgement} We thank Antony R. Lee, Tanja Fabsits, and Karishma Hathlia for a careful reading of the manuscript, useful comments, and fruitful discussions.

\appendix

\section{Derivation of the transformations in the phase-space formalism}\label{app:derivation_of_S_and_b}

Let us assume the most general Gaussian unitary from Eq.~\eqref{def:Gaussian_unitary}. Such Gaussian unitary transforms the vector of creation and annihilation operators from Eq.~\eqref{def:commutation_relation} as
\[\label{eq:Adash}
\boldsymbol{\hat{A}'}_{\!\!\!i}=\hat{U}^\dag \bA_i U.
\]
Because $\hat{U}$ is the Gaussian unitary, the transformation can be written as
\[\label{eq:Adash2}
\boldsymbol{\hat{A}'}=S\bA+\bb,
\]
where $S$ is the symplectic matrix satisfying Eqs.~\eqref{def:structure_of_S}. One can show that the first and the second moments transform according to the rule
\begin{subequations}\label{def:covariance_matrix_app}
\begin{align}
{\bd'}_{\!\!i}&:=\mathrm{tr}\big[{\hat{\rho}}\boldsymbol{\hat{A}'}_{\!\!\!i}\big]=(S\bd+\bb)_i,\\
{\sigma'}_{\!\!ij}&:=\mathrm{tr}\big[\hat{\rho}\,\{\Delta\boldsymbol{\hat{A}'}_{\!\!\!i},\Delta\boldsymbol{\hat{A}'}^\dag_{\!\!\!j}\}\big]=(S\sigma S^\dag)_{ij}.
\end{align}
\end{subequations}
The only question which remains to be answered is how the transformation depends on $W$ and $\bg$ from Eq.~\eqref{def:Gaussian_unitary}. In the following, we generalize the proof from~\cite{Luis1995a} which has been done so far only for $\bg=0$. We are going to use the identity
\[\label{eq:BHC_formula_corollary}
e^{\hat{X}}\bA_ie^{-\hat{X}}=\sum_{n=0}^\infty\frac{1}{n!}[\hat{X},\bA_i]_n,
\]
where $[\hat{X},\bA_i]_n=[\hat{X},[\hat{X},\bA_i]_{n-1}]$, $[\hat{X},\bA_i]_0=\bA_i$. Denoting $\hat{X}=-\tfrac{i}{2}\bA^\dag W \bA-\bA^\dag K \bg$, and using commutation relations
\[
[\hat{X},\bA_i]=(KW\bA)_i+\bg_i,
\]
we derive by induction
\[\label{eq:nth_element_of_BHC}
[\hat{X},\bA_i]_n=\big((iKW)^n\bA+(iKW)^{n-1}\bg\big)_i.
\]
Combining Eqs.~\eqref{eq:BHC_formula_corollary}, \eqref{eq:nth_element_of_BHC}, \eqref{eq:Adash}, and \eqref{eq:Adash2} yields
\begin{subequations}
\begin{align}
S&=e^{iKW},\\
\bb&=\sum_{n=0}^\infty\frac{(iKW)^n}{(n+1)!}\bg=\Big(\!\int_0^1e^{iKWt}\mathrm{d}t\!\Big)\ \!\bg.
\end{align}
\end{subequations}
For invertible $W$ we can also write
\[
\bb=(iKW)^{-1}\big(e^{iKW}-I\big)\bg.
\]

\section{List of the symplectic matrices in the complex and the real form formalism}\label{app:list_of_sympl_matrices}
To reduce the amount of confusion caused by different authors using different notations, we write what the symplectic matrices look like in the notation introduced by Eq.~\eqref{def:covariance_matrix}, and in one type of the so-called real form of the covariance matrix. Defining vectors of position and momenta operators $\boldsymbol{\hat{Q}}:=(\hat{x}_1,\dots,\hat{x}_N,\hat{p}_1,\dots,\hat{p}_N)^T$, where $\hat{x}_i:=\frac{1}{\sqrt{2}}(\hat{a}^\dag_i+\hat{a}_i)$, $\hat{p}_i:=\frac{i}{\sqrt{2}}(\hat{a}^\dag_i-\hat{a}_i)$, the real form displacement and the real form covariance matrix are defined as
\begin{subequations}\label{def:covariance_matrix_real}
\begin{align}
\bd_{{\mathrm{Re}} i}&=\mathrm{tr}\big[\hat{\rho}\boldsymbol{\hat{Q}}_i\big],\\
\sigma_{{\mathrm{Re}} ij}&=\mathrm{tr}\big[\hat{\rho}\,\{\Delta\boldsymbol{\hat{Q}}_i,\Delta\boldsymbol{\hat{Q}}_j\}\big],
\end{align}
\end{subequations}
where $\Delta\boldsymbol{\hat{Q}}:=\boldsymbol{\hat{Q}}-\bd_{{\mathrm{Re}}}$. The real form covariance matrix then transforms under the real form symplectic transformation as $\sigma_{\mathrm{Re}}'=S_{\mathrm{Re}}\sigma_{\mathrm{Re}} S^T_{\mathrm{Re}}$. Symplectic matrices in most other commonly used notations are simply obtained by rearranging some rows and columns of either the complex~\eqref{def:covariance_matrix} or the real form matrices~\eqref{def:covariance_matrix_real}. One-mode operations which leave the other modes invariant are easily lifted into two-mode operations by adding identities onto suitable places. For more information about the transformation between the real and the complex form see for example~\cite{Safranek2015b,Arvind1995a}.\\
\begin{widetext}
\noindent
\emph{Phase change (Rotation)} $\hat{R}(\theta)=\exp(-i\theta\hat{a}^\dag \hat{a})$, $\hat{R}_1(\theta)=\exp(-i\theta\hat{a}_1^\dag \hat{a}_1)$,
\[
R=\begin{bmatrix}
e^{-i\theta} & 0 \\
0 & e^{i\theta}
\end{bmatrix},
\quad R_{\mathrm{Re}}=\begin{bmatrix}
\cos\theta & \sin\theta \\
-\sin\theta & \cos\theta
\end{bmatrix},\quad \longrightarrow \quad R_1=\begin{bmatrix}
e^{-i\theta} & 0  & 0 & 0 \\
0 & 1 & 0 & 0 \\
0 & 0 & e^{i\theta} & 0 \\
0 & 0 & 0 & 1
\end{bmatrix},
\quad
R_{1{\mathrm{Re}}}=\begin{bmatrix}
\cos\theta & 0  & \sin\theta & 0 \\
0 & 1 & 0 & 0 \\
-\sin\theta & 0 & \cos\theta & 0 \\
0 & 0 & 0 & 1
\end{bmatrix}.
\]
\emph{One-mode squeezing} $\hat{S}(r,\chi)=\exp(-\frac{r}{2}(e^{i\chi}\hat{a}^{\dag2}-e^{-i\chi}\hat{a}^{2}))$,
\[
S=\begin{bmatrix}
\cosh r & -e^{i\chi}\sinh r \\
-e^{-i\chi}\sinh r & \cosh r
\end{bmatrix},
\quad S_{\mathrm{Re}}=\begin{bmatrix}
\cosh r-\cos\chi\sinh r & -\sin\chi\sinh r \\
-\sin\chi\sinh r & \cosh r+\cos\chi\sinh r
\end{bmatrix}.
\]
\emph{Mode-mixing} $\hat{B}(\theta,\chi)=\exp(\theta(e^{i\chi}\hat{a}_1^\dag\hat{a}_2-e^{-i\chi}\hat{a}_2^\dag\hat{a}_1))$,
\[
B=\begin{bmatrix}
\cos\theta & e^{i\chi}\sin\theta  & 0 & 0 \\
-e^{-i \chi}\sin\theta & \cos\theta & 0 & 0 \\
0 & 0 & \cos\theta & e^{-i \chi}\sin\theta \\
0 & 0 & -e^{i \chi}\sin\theta & \cos\theta
\end{bmatrix},
\quad B_{\mathrm{Re}}=\begin{bmatrix}
\cos\theta & \cos\chi\sin\theta & 0 & -\sin\chi\sin\theta \\
-\cos\chi\sin\theta & \cos\theta & -\sin\chi\sin\theta & 0 \\
0 & \sin\chi\sin\theta & \cos\theta & \cos\chi\sin\theta\\
\sin\chi\sin\theta & 0 & -\cos\chi\sin\theta & \cos\theta
\end{bmatrix}.
\]
\emph{Two-mode squeezing} $\hat{S}_{T}(r,\chi)=\exp(-r(e^{i\chi}\hat{a}_1^\dag\hat{a}_2^\dag-e^{-i\chi}\hat{a}_1\hat{a}_2))$,
\[
S_T\!=\!\begin{bmatrix}
\cosh r & 0  & 0 & -e^{i\chi}\sinh r \\
0 & \cosh r & -e^{i\chi}\sinh r & 0 \\
0 & -e^{-i\chi}\sinh r & \cosh r & 0 \\
-e^{-i\chi}\sinh r & 0 & 0 & \cosh r
\end{bmatrix},\ S_{T{\mathrm{Re}}}\!=\!\begin{bmatrix}
\cosh r & -\cos \chi\sinh r  & 0 & -\sin \chi\sinh r \\
-\cos \chi\sinh r & \cosh r & -\sin \chi\sinh r & 0 \\
0 & -\sin \chi\sinh r & \cosh r & \cos \chi\sinh r \\
-\sin \chi\sinh r & 0 & \cos \chi\sinh r & \cosh r
\end{bmatrix}.
\]

\section{Full expressions for two-mode squeezing and mode-mixing channels}\label{app:full_expressions}
Using the probe state from Eq.~\eqref{eq:simplified_probe_state}, and defining $\phi_\chi:=\phi_1+\phi_2+\chi$, $\phi_{1\chi}:=\phi_1-\phi_{d2}+\chi$, $\phi_{2\chi}:=\phi_2-\phi_{d1}+\chi$, the quantum Fisher information for the estimation of a two-mode squeezing channel $\hat{S}_{T}(\epsilon,\chi)$ reads
\[
\begin{split}
&H(\epsilon)=\\
&2\cos^2(2\theta)\bigg(\!\frac{(\lambda_1\!+\!\lambda_2)^2}{\lambda_1\lambda_2\!+\!1}\big(\!\cos^2\!\!\phi_\chi\cosh^2(\!r_1\!-\!r_2\!)\!+\!\sin^2\!\!\phi_\chi\cosh^2(\!r_1\!+\!r_2\!)\big)
+\frac{(\lambda_1\!-\!\lambda_2)^2}{\lambda_1\lambda_2\!-\!1}\big(\!\cos^2\!\!\phi_\chi\sinh^2(\!r_1\!-\!r_2\!)\!+\!\sin^2\!\!\phi_\chi\sinh^2(\!r_1\!+\!r_2\!)\!\big)\!\bigg)\\
&+4\sin^2(2\theta)\bigg(\frac{\lambda_1^2}{\lambda_1^2+1}\big(\cos^2(\phi_\chi+2\psi)+\sin^2(\phi_\chi+2\psi)\cosh(2r_1)\big)
+\frac{\lambda_2^2}{\lambda_2^2+1}\big(\cos^2(\phi_\chi-2\psi)+\sin^2(\phi_\chi-2\psi)\cosh(2r_2)\big)\bigg)\\
&+\frac{4}{\lambda_1}\Big(e^{2r_1}\big(\norm{{d}_1}\sin\theta\cos(\phi_{2\chi}+\psi)-\norm{{d}_2}\cos\theta\cos(\phi_{1\chi}+\psi)\big)^2
+e^{-2r_1}\big(\norm{{d}_1}\sin\theta\sin(\phi_{2\chi}+\psi)-\norm{{d}_2}\cos\theta\sin(\phi_{1\chi}+\psi)\big)^2\Big)\\
&+\frac{4}{\lambda_2}\Big(e^{2r_2}\big(\norm{{d}_1}\sin\theta\cos(\phi_{2\chi}-\psi)+\norm{{d}_2}\cos\theta\cos(\phi_{1\chi}-\psi)\big)^2
+e^{-2r_2}\big(\norm{{d}_1}\sin\theta\sin(\phi_{2\chi}-\psi)+\norm{{d}_2}\cos\theta\sin(\phi_{1\chi}-\psi)\big)^2\Big).\\
\end{split}
\]
Using the probe state from Eq.~\eqref{eq:simplified_probe_state}, and defining $\phi_\chi:=\phi_1-\phi_2+\chi$, $\phi_{1\chi}:=\phi_1+\phi_{d2}+\chi$, $\phi_{2\chi}:=\phi_2+\phi_{d1}-\chi$, the quantum Fisher information for the estimation of a mode-mixing channel $\hat{B}(\epsilon,\chi)$ reads
\[
\begin{split}
&H(\epsilon)=4\sin^2(2\theta)\sin^2\!\!\phi_\chi\bigg(\frac{\lambda_1^2}{\lambda_1^2+1}\sinh^2(2r_1)
+\frac{\lambda_2^2}{\lambda_2^2+1}\sinh^2(2r_2)\bigg)\\
&\!\!+\!\frac{2(\lambda_1\!+\!\lambda_2)^2}{\lambda_1\lambda_2\!+\!1}\!\Big(\!\big(\!\cos(\!2\theta\!)\sin\!\phi_\chi\sin(\!2\psi\!)\!-\!\cos\!\phi_\chi\cos(\!2\psi\!)\!\big)^2\!\!\sinh^2(\!r_1\!-\!r_2\!)
\!+\!\big(\!\cos(\!2\theta\!)\sin\!\phi_\chi\cos(\!2\psi\!)\!+\!\cos\!\phi_\chi\sin(\!2\psi\!)\!\big)^2\!\!\sinh^2(\!r_1\!+\!r_2\!)\!\Big)\\
&\!\!+\!\frac{2(\lambda_1\!-\!\lambda_2)^2}{\lambda_1\lambda_2\!-\!1}\!\Big(\!\big(\!\cos(\!2\theta\!)\sin\!\phi_\chi\sin(\!2\psi\!)\!+\!\cos\!\phi_\chi\cos(\!2\psi\!)\!\big)^2\!\!\cosh^2(\!r_1\!-\!r_2\!)
\!+\!\big(\!\cos(\!2\theta\!)\sin\!\phi_\chi\cos(\!2\psi\!)\!+\!\cos\!\phi_\chi\sin(\!2\psi\!)\!\big)^2\!\!\sinh^2(\!r_1\!+\!r_2\!)\\
&\ \ \ \ \ \ \ \ \ \ \ \ \ \ \ \ +\frac{1}{2}\cos(\!2\theta\!)\sin(\!2\phi_\chi\!)\sin(\!4\psi\!)\sinh(2r_1)\sinh(2r_2)\!\Big)\\
&\!\!+\frac{4}{\lambda_1}\Big(e^{2r_1}\big(\norm{{d}_1}\sin\theta\cos(\phi_{2\chi}+\psi)+\norm{{d}_2}\cos\theta\cos(\phi_{1\chi}+\psi)\big)^2
+e^{-2r_1}\big(\norm{{d}_1}\sin\theta\sin(\phi_{2\chi}+\psi)+\norm{{d}_2}\cos\theta\sin(\phi_{1\chi}+\psi)\big)^2\Big)\\
&\!\!+\frac{4}{\lambda_2}\Big(e^{2r_2}\big(\norm{{d}_1}\cos\theta\cos(\phi_{2\chi}-\psi)-\norm{{d}_2}\sin\theta\cos(\phi_{1\chi}-\psi)\big)^2
+e^{-2r_2}\big(\norm{{d}_1}\cos\theta\sin(\phi_{2\chi}-\psi)-\norm{{d}_2}\sin\theta\sin(\phi_{1\chi}-\psi)\big)^2\Big).\\
\end{split}
\]
\end{widetext}

\bibliographystyle{apsrev}
\bibliography{Optimal_probe_states_BiBTeX}

\end{document}